%% file: main.tex
\documentclass[runningheads]{llncs}
\usepackage[T1]{fontenc}

\usepackage[ruled,vlined]{algorithm2e}
\usepackage{hyperref}
\usepackage{stmaryrd}

\usepackage{times}
\usepackage{soul}
\usepackage[small]{caption}
\usepackage{graphicx}
\usepackage{amsmath}
\usepackage{amssymb}
\usepackage{booktabs}
\usepackage{color}
\usepackage[misc]{ifsym}

\begin{document}
\title{DistSPECTRL: Distributing Specifications in Multi-Agent Reinforcement Learning Systems}
\titlerunning{DistSPECTRL: Distributing Specs. in MARL Systems}
\author{Joe~Eappen (\Letter) \and Suresh Jagannathan}
%

\authorrunning{J. Eappen and S. Jagannathan}
%
%
\institute{Purdue University, West Lafayette IN, 47907, USA 
	\\
	\email{jeappen@purdue.edu}\\
\email{suresh@cs.purdue.edu}}


\input{sections/macros}

\toctitle{DistSPECTRL: Distributing Specifications in Multi-Agent Reinforcement Learning Systems}
\tocauthor{Joe~Eappen (Purdue), Suresh~Jagannathan (Purdue)}
\maketitle

	\begin{abstract}
		
		While notable progress has been made in specifying and learning objectives for
		general cyber-physical systems, applying these methods to distributed
		multi-agent systems still pose significant challenges.  Among these are the need
		to (a) craft specification primitives that allow expression and
		interplay of both local and global objectives, (b) tame explosion in
		the state and action spaces to enable effective learning, and (c)
		minimize coordination frequency and the set of engaged participants
		for global objectives.  To address these challenges, we propose a
		novel specification framework that allows natural composition of local
		and global objectives used to guide training of a multi-agent system.
		Our technique enables learning expressive policies that allow agents
		to operate in a coordination-free manner for local objectives, while using a
		decentralized communication protocol for enforcing global ones.
		Experimental results support our claim that sophisticated multi-agent
		distributed planning problems can be effectively realized using
		specification-guided learning. 
		Code is provided at 
		\url{https://github.com/yokian/distspectrl}.

	\end{abstract}

	\section{Introduction}
	
	Reinforcement Learning (RL) can be used to learn complex
        behaviors in many different problem settings.  A main
        component of RL is providing feedback to an agent via a reward
        signal. This signal should encourage desired behaviors, and
        penalize undesirable ones, enabling the agent to eventually
        proceed through a sequence of tasks and is designed by the
        programmer beforehand. A commonly used technique to encode
        tasks in a reward signal is the sparse method of providing
        zero reward until a task is completed upon which a non-zero
        reward is given to the agent.  Because this procedure has the
        significant shortcoming of delaying generating a useful
        feedback signal for a large portion of the agent-environment
        interaction process, a number of alternative techniques have
        been proposed \cite{Ng99policyinvariance,NIPS2017_453fadbd}.
        
        Formulating a reward signal that reduces the sparsity of this
        feedback is known as \emph{reward shaping}.  Often, this is
        done manually but a more general, robust method would be to
        automatically shape a reward given a specification of desired
        behavior.  \spectrl\cite{NIPS2019_9462} proposes a reward
        shaping mechanism for a set of temporal logic specifications
        on a single-agent task that uses a compiled a finite-state
        automaton called a task monitor.  Reward machines
        \cite{tor-etal-arxiv20,cam-etal-ijcai19,tor-etal-icml18} are
        another objective-specifying method for RL problems that also
        define a finite automaton akin to the ones used in \spectrl,
        with some subtle differences such as the lack of registers
        (used by the task monitor for memory).
	
	Distributed multi-agent applications however, introduce new
        challenges in automating this reward shaping process.  Agents
        have their own respective goals to fulfill as well as
        coordinated goals that must be performed in cooperation with
        other agents. While the expressiveness of the language in
        \spectrl lends itself, with minor extensions, to specifying
        these kinds of goals, we require new compilation and execution
        algorithms to tackle inherent difficulties in multi-agent
        reinforcement learning (MARL); these include credit assignment
        of global objectives and the presence of large state and
        action spaces that grow as the number of agents increases.
        Learning algorithms for multi-agent problems have often
        encouraged distribution as a means of scaling in the presence
        of state and action space explosion.  This is because purely
        centralized approaches have the disadvantages of not only
        requiring global knowledge of the system at all times but also
        induce frequent and costly synchronized agent control.
        	
	To address these issues, we develop a new specification-guided
        distributed multi-agent reinforcement learning framework.  Our
        approach  has four main features.  First, we introduce two
        classes of predicates (\textit{viz.} local and global) to
        capture tasks in a multi-agent world
        (Sec.~\ref{sec:spec_maw}). Second, we develop a new procedure
        for generating composite task monitors using these predicates
        and devise new techniques to distribute these monitors over
        all agents to address scalability and decentralization
        concerns (Sec.~\ref{ssec:create_tm}).  Third, we efficiently
        solve the introduced problem of subtask synchronization
        (Sec.~\ref{ssec:task_synch}) among agents via synchronization
        states in the task monitors.  Lastly, we describe a wide class
        of specification structures (Sec. \ref{sec:ma_spec_prop})
        amenable to scaling in the number of agents and provide a
        means to perform such a scaling
        (Sec. \ref{sec:scaling_method}).
        
	
	By using these components in tandem, we provide the first
        solution to composing specifications and distributing them
        among agents in a scalable fashion within a multi-agent
        learning scenario supporting continuous state and action spaces.
        Before presenting details of our approach, we first provide
        necessary background information (Sec.~\ref{sec:background})
        and formalize the problem (Sec.~\ref{sec:problem}).
        
	
	\section{Background}
        \label{sec:background}
	
	\subsubsection{\textbf{Markov Decision Processes}}
	Reinforcement learning is a tool to solve Markov Decision
        Processes (MDPs). MDPs are tuples of the form $\langle \mathcal{S},D,A,P,R,T
        \rangle$ where $\mathcal{S}\in \mathbb{R}^n$ is the state
        space, $D$ is the initial state distribution, $A\in
        \mathbb{R}^m$ is the action space, $P: \mathcal{S}\times
        A\times \mathcal{S} \rightarrow [0,1]$ is the transition
        function, and $T$ is the time horizon. A rollout $\zeta\in Z$
        of length $T$ is a sequence of states and actions $\zeta =
        (s_0,a_0,...,a_{T-1},s_T)$ where $s_i\in \mathcal{S}$ and
        $a_i\in A$ are such that $s_{i+1} \sim
        P(s_i,a_i)$. $R:Z\rightarrow \mathbb{R}$ is a reward
        function used to score a rollout $\zeta$.
	
		\subsubsection{\textbf{Multi-Agent Reinforcement Learning}}
	A Markov game with $\mathcal{N}=\{1,\cdots,N\}$
	denoting the set of $N$ agents is a tuple 
	$\mathcal{M}_g=\langle
        \mathcal{N},\{\mathcal{S}^i\}_{i\in
          \mathcal{N}},D,\{A^i\}_{i\in \mathcal{N}},P,\{R^i\}_{i\in
          \mathcal{N}},T \rangle$
       where ${A^i}$, ${R^i}$ define their
        agent-specific action spaces and reward functions.  They are a
        direct generalization of MDPs to the multi-agent scenario. Let
        $\mathcal{S}_m=\{\mathcal{S}^i\}_{i\in \mathcal{N}}$ and
        $A_m=\{A^i\}_{i\in \mathcal{N}}$, then $P:\mathcal{S}_m\times
        A_m \times \mathcal{S}_m \rightarrow [0,1]$ is the transition
        function. A rollout $\zeta_m\in Z_m$ here corresponds to
        $\zeta_m = ( \bar{s}_0,\bar{a}_0,...,\bar{a}_{T-1},\bar{s}_T)$
        where $\bar{s}\in \mathcal{S}_m$ and $\bar{a}\in A_m$. We 
        also define an agent specific rollout $\zeta^i_m\in
        Z^i_m,~\zeta^i_m = ( s^i_0,a^i_0,...,a^i_{T-1},s^i_T)$ where
        $s^i\in \mathcal{S}^i$ and $a^i\in A^i$. $D$ is the initial
        state distribution over $\mathcal{S}_m$.
	
	Agents attempt to learn a policy $\pi^i : \mathcal{S}^i \rightarrow
        \Delta(A^i)$ such that $\mathbb{E}\left[ \sum_t R^i_t | \pi^i,
          \pi^{-i} \right]$ is maximized, where $\Delta(A^i)$ is a
        probability distribution over $A^i$ and $\pi^{-i}$ is the set
        of all policies apart from $\pi^{i}$. We use $\Pi=\{\pi^i\}_{i
          \in \mathcal{N} }$ to denote the set of all agent policies.
	For simplicity, we restrict our formulation to a
        \textit{homogeneous} set of agents which operate over the same state
        ($\mathcal{S}^i=\mathcal{S}_A$) and action space ($A^i=A_A$).
	
		\subsubsection{\textbf{SPECTRL}}
		Jothimurugan \emph{et. al}~\cite{NIPS2019_9462}
                introduce a specification language for reinforcement
                learning problems built using temporal logic
                constraints and predicates. It is shown to be adept at
                handling complex compositions of task specifications
                through the use of a \emph{task monitor} and
                well-defined monitor transition rules.  Notably, one
                can encode Non-Markovian tasks into the MDP using the
                additional states of the automaton (task monitor)
                compiled from the given specification.
	
	The atomic elements of this language are Boolean predicates
        $b$ defined as functions of a state $\mathcal{S}$ \edit{with output $\pred{b}: \mathcal{S} \rightarrow \mathbb{B}$}. These
        elements have quantitative semantics $\quantpred{b}$ with the
        relation being $\pred{b}(s)=\mathtt{True} \iff \quantpred{b} (s) >
        0$. Specifications $\phi$ are Boolean functions of the state
        trajectory $\zeta=(s_1,s_2,...,s_T)$. The specification
        language also includes composition functions for a
        specification $\phi$ and Boolean predicate $b$, with the
        language defined as
	$$\phi :: = \mathtt{achieve}~b ~|~ \phi~\mathtt{ensuring }~b ~|~ \phi_1 ;~ \phi_2 ~|~ \phi_1~ \mathtt{or}~ \phi_2$$
	The description of these functions is as
        follows. $\mathtt{achieve}~b$ is true when the trajectory 
        satisfies $b$ at least once. $\phi~\mathtt{ensuring }~b$ is
        true when $b$ is satisfied at all timesteps in the
        trajectory. $\phi_1 ;~ \phi_2$ is a sequential operator that
        is true when, in a given trajectory $\zeta=(s_1,s_2,...,s_T)$,
        $\exists~k >1$ such that $\phi_1(s_1,...,s_k)$ is true and
        $\phi_2(s_{k+1},...,s_T)$ is true. In other words, $\phi_1 ;~
        \phi_2$ represents an ordered sequential completion of
        specification $\phi_1$ followed by $\phi_2$.  Lastly, $\phi_1~
        \mathtt{or}~ \phi_2$ is true when a trajectory satisfies
        either $\phi_1$ or $\phi_2$.
	
	Given a specification $\phi$ on a Markov Decision Process
        $\langle S,D,A,P,T \rangle$ (MDP) defined using \spectrl,
        a task monitor $\langle Q, X, \Sigma,
        U , \Delta, q_0, v_0, F, \rho \rangle$ (a finite state
        automaton \cite{VARDI19941}) is compiled to record the completion status
        of tasks with monitor states $Q$; final monitor states $F$
        denote a satisfied trajectory. This is used to create an
        augmented version of the MDP $\langle
        \tilde{S},\tilde{s_0},\tilde{A},\tilde{P},\tilde{R},\tilde{T}
        \rangle$ with an expanded state, action space and modified
        reward function . The task monitor provides a scoring function
        for trajectories in the augmented MDP to guide policy
        behavior.
	
	While \spectrl has been shown to work with trajectory-based
        algorithms for reinforcement learning
        \cite{mania_simple_2018}, it is not immediately evident how to
        translate it to common RL algorithms such as DDPG
        \cite{silver_deterministic_nodate} and
        PPO~\cite{DBLP:journals/corr/SchulmanWDRK17}.  A simple
        solution would be to keep the episodic format with a
        trajectory $\zeta=(s_0,\cdots,s_T)$ and assign the trajectory
        value of \spectrl (a function of $\zeta$) to the final state
        transition in the trajectory $s_{T-1}\rightarrow s_T$ and zero
        for all other states. 
        \edit{Importantly}, this maintains the trajectory ordering
        properties of \spectrl in the episodic return $(\sum_{t=0}^T
        r_t)$.
	
	\section{Problem Statement}
        \label{sec:problem}

	Directly appropriating \spectrl for our use case of imposing
        specifications on multi-agent problems poses significant
        scalability issues. Consider the case
	$$\phi_a = \mathtt{achieve}(\mathtt{reach}(P));\mathtt{achieve}(\mathtt{reach}(Q))$$
	
	\noindent \edit{ where $\pred{\mathtt{reach}(P)}=\mathtt{True}$ when an agent reaches state $P$. }
	 To ease the illustration of our framework, we assume
        that all agents are homogeneous, \textit{i.e.} $\mathcal{S}^i
        = \mathcal{S}_A, \forall i \in \mathcal{N}$. Now, the state
        space of the entire multi-agent system is $\mathcal{S}
        =(\mathcal{S}_A)^N$ for $N$ agents (we omit $m$ for perspicuity).
	
	If the predicate $\mathtt{reach}$ was defined on the entire
        state $\mathcal{S}$, it would yield a specification forcing
        synchronization between agents. On the other hand, if
        $\mathtt{reach}$ was defined on the agent state
        $\mathcal{S}_A$, then it would create a \textit{localized}
        specification where synchronization is not required. This
        would be akin to allowing individual agents to act
        independently of other agent behaviors.

	However, using a centralized task monitor for the localized
        predicate would cause the number of monitor states to
        exponentially increase with the number of agents $N$ and
        subtasks $K$ since the possible stages of task completion
        would be $\mathcal{O}(K^N)$.
	
	To address this scalability issue, the benefits of task
        monitor distribution are apparent. In the case of $\phi_a$
        above, assume $\mathtt{reach}$ is defined on the local state
        space $\mathcal{S}_A$. If each agent had a separate task
        monitor stored locally to keep track of the task completion
        stages, the new number of monitor states is now reduced to
        $\mathcal{O}(NK)$.
        
        \edit{
        Consider an example of robots in a warehouse.
        A few times a day, all
        robots must gather at a common point for damage inspection at the same time (akin 
        to a global reach) to minimize the frequency of 
        inspection (an associated cost). To ensure satisfaction of the entire specification, 
        the reward given to an RL agent learning this objective should capture both the 
        global and local tasks. For example, if the global reach task for the routine 
        inspection is made local instead, the cost incurred may be larger than if 
        it was a synchronized global objective.}
	
	\subsubsection*{\textbf{Main Objective}} Given a specification $\phi$
        on a system of $N$ agents, we wish to find policies
        $\Pi=\{\pi^1,\cdots,\pi^N\}$ to maximize the probability of
        satisfying $\phi$ for all agents.  Formally, we seek
	$$ \Pi^* \in
        \underset{\pi^1,\cdots,\pi^N}{\arg\max}~~\underset{\zeta_m
          \sim D_\Pi}{\text{Pr}} \left[ \pred{\phi(\zeta_m)} = \mathtt{True}
          \right] $$
	\noindent where $D_\Pi$ is the distribution of all system
        rollouts when all agents collectively follow policy set $\Pi$.
        We emphasize that $\phi$ acts on the entire rollout,
        $\phi:Z_m\rightarrow \{0,1\}$ and not in an agent-specific
        manner, $\phi':Z^i_m\rightarrow \{0,1\}$. This discourages
        agents from attempting to simply satisfy their local
        objectives while preventing the system from achieving
        necessary global ones.

	\label{sec:ps}
	
	\begin{figure}[!h]
		\begin{center}
			\includegraphics[width=0.7\linewidth]{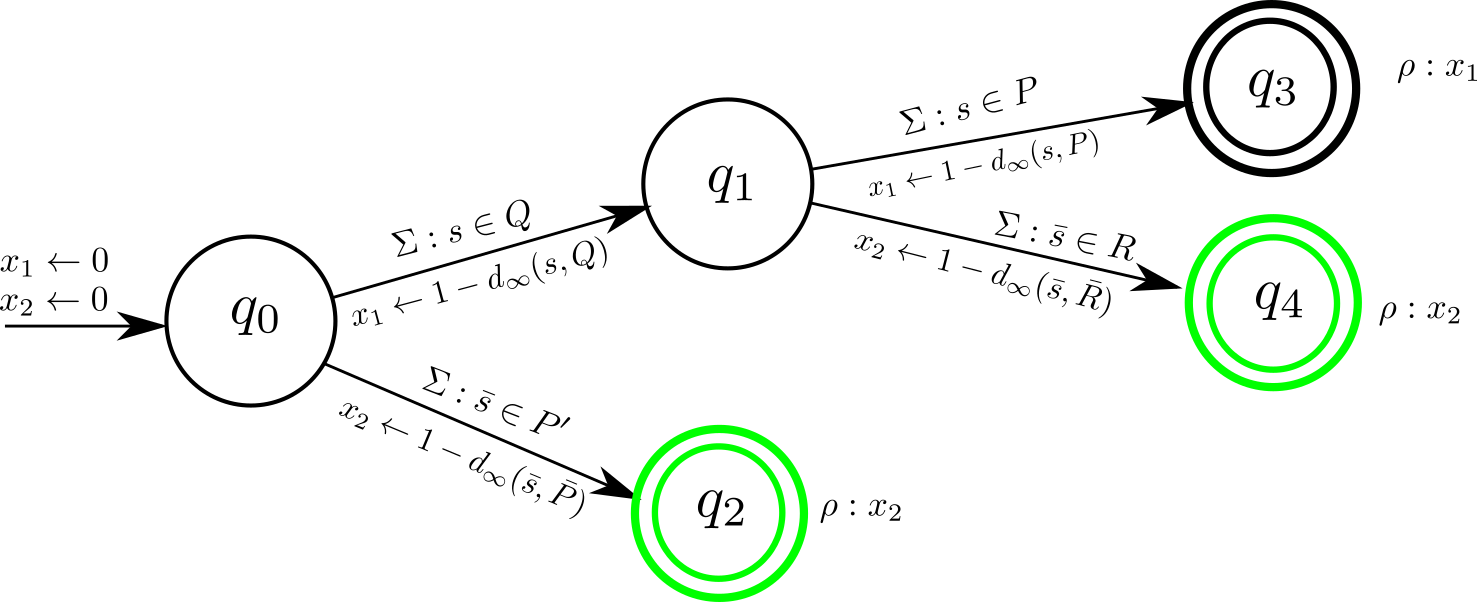}
		\end{center}
		\caption{Example Composite Task Monitor for
                  specification $\phi_{ex}$
                  (Sec.~\ref{ssec:tm_example}) with 4 task goals
                  denoted by Q,P,P' and R where the agent starts at
                  $q_0$. Double circles represent \emph{final} states
                  while green circles represent \emph{global}
                  states. The diagram removes a state between $q_1$
                  and $(q_3,q_4)$ as well as self-loops for ease of
                  explanation.}
		\label{fig:example_spec}
	\end{figure}

	\section{SPECTRL in a Multi-Agent World}
	\label{sec:spec_maw}
	Unlike the single agent case, multi-agent problems have two
        major classes of objectives. Agents have individual goals to
        fulfill as well as collective goals that require coordination
        and/or global system knowledge.  These individual goals are
        often only dependent on the agent-specific state $s^i$ while
        collective goals require full system knowledge $\bar{s}$.
	
	Consequently, for a multi-agent problem, we see the need for
        two types of predicates \textit{viz.} local and global. Local
        predicates are of the form $p_{lo}:\mathcal{S}_A\rightarrow
        \mathbb{B}$ whereas global predicates have the form
        $p_{gl}:\mathcal{S}\rightarrow \mathbb{B}$ where $\mathbb{B}$
        is the Boolean space. We introduce two simple extensions of
        $\mathtt{reach}$~\cite{NIPS2019_9462} to demonstrate the
        capabilities of this distinction.

	Local predicates are defined with respect to each agent and
        represent our individual goals. As an example, closely related
        to the problems observed in \spectrl, we introduce the following
        local predicates for a state $s_a \in \mathcal{S}_A$,
	$$\pred{\mathtt{reach}_{lo}x}(s_a) = (||s_a-x||_\infty < 1)$$
	\noindent which represents reaching near location $x$ in terms of the $L_\infty$ norm. 
	Now to enforce global restrictions, we introduce counterparts
        to these predicates that act on a global state $\bar{s} \in
        \mathcal{S}$.
	$$\pred{\mathtt{reach}_{gl}\bar{x}}(\bar{s}) = (||\bar{s}-\bar{x}||_\infty < 1)$$
	\noindent where we now have a set of locations $\bar{x}\in \mathcal{S}$. 
	
	As in \spectrl, each of these predicates $b$ require
        quantitative semantics $\quantpred{b}$ to facilitate our
        reward shaping procedure. We define these semantics as follows:
	\begin{itemize}
		\item  $\mathtt{reach}_{lo}$ 
		has the same semantics as  $\mathtt{reach}$ 
		in \cite{NIPS2019_9462} yet is defined on space $\mathcal{S}_A$.
		$$\quantpred{\mathtt{reach}_{lo}x}(s_a) = 1-d_\infty(s_a,x)$$
		\noindent where $d_\infty(a,b)$ represents the
                $L_\infty$ distance between $a$ and $b$ with the usual
                extension to the case where $b$ is a set.

		\item $\mathtt{reach}_{gl}$ is defined on the state space $\mathcal{S}$ as
		$$\quantpred{\mathtt{reach}_{gl}\bar{x}}(\bar{s}) = 1-d_\infty(\bar{s},\bar{x})$$
	\end{itemize}
	
	We observe that the same composition rules can apply to these
        predicates and we thus attempt to solve RL systems described with
        these compositions.  As shown in Sec.~\ref{sec:ps}, using a
        centralized \spectrl compilation algorithm on the entire state
        space, even for simple sequences of tasks, leads to an
        explosion in monitor states.  We, therefore, distribute task
        monitors over agents to handle scalability. Furthermore, we
        also need to change \spectrl's compilation rules to handle
        mixed objective compositions such as\footnote{We omit
          $\mathtt{achieve}$ in
          $\mathtt{achieve}(\mathtt{reach}_{lo}(P))$ and
          $\mathtt{achieve}(\mathtt{reach}_{gl}(P))$ from here on to
          reduce clutter; this specification is implied when we compose
          $\mathtt{reach}(P)$ with $;$ and $\mathtt{or}$.} $$\phi=
        \mathtt{reach}_{lo}(P);
        \mathtt{reach}_{gl}(Q);\mathtt{reach}_{gl}(R)$$
	
	To compile these specifications into a usable format, we
        utilize a \textit{composite task monitor} as described in
        Sec.~\ref{ssec:create_tm} and develop a new algorithm to achieve
        our goal.
	\label{ssec:tm_example}
	As an example, see Fig. \ref{fig:example_spec} depicting a task monitor whose specification is:
	\begin{align*}
	\phi_{ex}=&  \mathtt{reach}_{gl}(P')\ \mathtt{or}\ \mathtt{reach}_{lo}(Q) ; \left[ \mathtt{reach}_{lo}(P) \ \mathtt{or}\  \mathtt{reach}_{gl}(R)   \right]
	\end{align*}%
	Here, we have 4 task goals denoted by $P,Q,R$ and $P'$. The
        agents all start at the root node $q_0$. States $q_2$, $q_3$ and $q_4$
        are all final states in the task monitor while $q_2$ and
        $q_4$ are global monitor states. As shown in Sec.~\ref{ssec:id_synch},
        $q_0$ and $q_1$ are a \textit{synchronization states}. While
        it may seem that agents only require coordination at global
        states, it is also necessary for the agents to have the same
        task transition at these synchronization states as well.

	\section{Compilation Steps}

        Given a specification $\phi$ and the Markov game
        $\mathcal{M}_g$, we create a task monitor $M$ that is
        distributed among agents by making agent-specific copies. This
        is used to create an augmented Markov game $\mathcal{M}'_g = \langle \mathcal{N},\{\tilde{\mathcal{S}}_A\}_{i\in
        	\mathcal{N}},\tilde{D},\{\tilde{A}_A\}_{i\in
        	\mathcal{N}},\tilde{P},\{\tilde{R}^i\}_{i\in \mathcal{N}},T
        \rangle$ on which the individual agent policies are trained.

	\subsubsection{\textbf{Create Composite Task Monitor}}
	\label{ssec:create_tm}

	When the types of specifications are divided into two based on
        the domain, the solution can be modeled with a
        \textit{composite task monitor} $M_\phi = \langle Q, \tilde{X},
        \tilde{\Sigma}, \tilde{U} , \tilde{\Delta}, q_0, v_0, F, \rho
        \rangle$.  As in \spectrl, $Q$ is a finite set of monitor
        states. $\tilde{X} = X_l\cup X_g$ is a finite set of registers
        that are partitioned into $X_l$ for local predicates and $X_g$
        for global predicates.  These registers are used to keep track
        of the degree of completion of the task at the current monitor
        state for local and global tasks respectively.
	
	We describe below how to use the compiled composite task
        monitor to create an augmented Markov game
        $\mathcal{M}_g'$. Each $\tilde{\mathcal{S}}_A$ in
        $\mathcal{M}_g'$ is an augmented state space with an augmented
        state being a tuple $(s_A,q,v)\in \mathcal{S}_A\times Q\times
        V $ where $V\in \mathbb{R}^X$ and $v\in V$ is a vector
        describing the register values.
	
	$\tilde{\Delta}=\Delta_l \cup \Delta_g$  houses the
        transitions of our task monitor. We require that: i) different
        transitions are allowed only under certain conditions
        defined by our states and register values; and, ii) furthermore,
        they must also provide rules on how to update the
        register values during each transition.  To define these
        conditions for transition availability, we use
        $\tilde{\Sigma}=\Sigma_l\cup\Sigma_g$ where $\Sigma_l$ is a
        set of predicates over $\mathcal{S}_A\times V$ and $\Sigma_g$
        is a set of predicates over $\mathcal{S}\times V$. Similarly,
        $\tilde{U}=U_l\cup U_g$ where $U_l$ is a set of functions
        $u_l:\mathcal{S}_A\times V\rightarrow V$ and $U_g$ is a set of
        functions $u_g:\mathcal{S}\times V\rightarrow V$.  Now, we can
        define $\tilde{\Delta}\subseteq Q\times \tilde{\Sigma}\times
        \tilde{U}\times Q$ to be a finite set of transitions that are
        non-deterministic. Transition $(q,\sigma,u,q')\in \tilde{\Delta}$ is an
        augmented transition either representing
        $(s^i,q,v)\xrightarrow{a^i|\Pi_{-i}}((s^{i})',q',u_l(s^i,v))$
        or the form
        $(\bar{s},q,v)\xrightarrow{a^i|\Pi_{-i}}(\bar{s}',q',u_g(s,v))$
        depending on whether $\sigma\in \Sigma_l$ or $\sigma\in
        \Sigma_g$ respectively. Let $\delta_l\in \Delta_l$ represent
        the former (localized) and $\delta_g \in \Delta_g$ the latter
        (global) transition types.  Here $\Pi_{-i}$ denotes the policy
        set of all agents except agent $i$. Lastly, $q_0$ is the
        initial monitor state and $v_0$ is the initial register value
        (for all agents), $F\subseteq Q$ is the set of final monitor
        states, and $\rho:\mathcal{S}\times F \times V\rightarrow
        \mathbb{R}$ is the reward function. 
	
	Copies of these composite task monitors $M$ are distributed
        over agents $\mathcal{N}$ to form the set
        $\{M^i\}_{i\in\mathcal{N}}$. These individually stored task
        monitors are used to let each agent $i\in \mathcal{N}$ keep
        track of its subtasks and the degree of completion of those
        subtasks by means of monitor state $q^i$ and register value
        $v^i$.
	
	\subsubsection{\textbf{Create Augmented Markov Game}}
	\label{ssec:crt_mkv_game}

 	From our specification 	$\phi$ we create the augmented Markov
        game 
        $\mathcal{M}_g' = \langle
        \mathcal{N},\{\tilde{\mathcal{S}}_A\}_{i\in
          \mathcal{N}},\tilde{D},\{\tilde{A}_A\}_{i\in
          \mathcal{N}},\tilde{P},\{\tilde{R}^i\}_{i\in \mathcal{N}},T
        \rangle$ using the compiled composite task monitor $M$
      . A
        set of policies $\tilde{\Pi}^*$ that maximizes rewards in
        $\mathcal{M}_g'$ should maximize the chance of the
        specification $\phi$ being satisfied.
	
	Each $\tilde{\mathcal{S}}_A = \mathcal{S}_A\times Q\times V$
        and $\tilde{D}=(\{s_0\}_{i\in\mathcal{N}},q_0,v_0)$. We use
        $\Delta$ to augment the transitions of $P$ with monitor
        transition information. Since $\Delta$ may contain
        non-deterministic transitions, we require the policies
        $\tilde{\Pi}$ to decide which transition to choose. Thus
        $\tilde{A}_A= A_A\times A_\phi $ where $A_\phi=\Delta$ 
        chooses among the set of available transitions at a monitor
        state $q$.  Since monitors are distributed among all
        agents in $\mathcal{N}$, we denote the set of current monitor
        states as $\bar{q}=\{q^i\}_{i\in\mathcal{N}}$ and the set of
        register values as $\bar{v}=\{v^i\}_{i\in\mathcal{N}}$.  Now,
        each agent policy must output an \textit{augmented action}
        $(a,\delta)\in \tilde{A}_A $ with the condition that $\delta_l
        = (q,\sigma_l,u_l,q')$ is possible in local augmented state
        $\tilde{s}_a=(s_a,q,v)$ if $\sigma_l(s_a,v)$ is
        $\mathtt{True}$ and $\delta_g = (q,\sigma_g,u_g,q')$ is
        possible in global augmented state
        $\tilde{s}=(\bar{s},\bar{q},\bar{v})$ if $\sigma_g(\bar{s},v)$
        is $\mathtt{True}$.  We can write the augmented transition
        probability $\tilde{P}$ as,
	$$ \tilde{P}((\bar{s},q,v),(a,(q,\sigma,u,q')),(\bar{s}',q',u(\bar{s},v)))) = P(\bar{s},a,\bar{s}') $$ for transitions $\delta_g\in\Delta_g$ with $(\sigma,u)=(\sigma_g,u_g)$ and transitions $\delta_l\in\Delta_l$ with $(\sigma,u)=(\sigma_l,u_l)$. 
	Here, we let $u_l(\bar{s},v) = u_l(s^i,v)$ for agent $i$ since $s^i$ is included in $\bar{s}$.	
	An \textit{ augmented rollout}  $\tilde{\zeta}_m$ where
	$$\tilde{\zeta}_m = (
        (\bar{s}_0,\bar{q}_0,\bar{v}_0),\bar{a}_0,...,\bar{a}_{T-1},(\bar{s}_T,\bar{q}_T,\bar{v}_T)
        )$$ is formed by these augmented transitions. To translate this
        trajectory back into the Markov game $\mathcal{M}_g$ we can
        perform projection $\project(\tilde{\zeta}_m) =
        (\bar{s}_0,\bar{a}_0,...,\bar{a}_{T-1},\bar{s}_T, ) $.
	
	\subsubsection{\textbf{Determine Shaped Rewards}}

	Now that we have the augmented Markov game $\mathcal{M}_g'$
        and compiled our composite task monitor, we proceed to form
        our reward function that encourages the set of policies $\Pi$
        to satisfy our specification $\phi$.  We can perform shaping
        in a manner similar to \spectrl's single-agent case on our
        distributed task monitor. Crucially, since reward shaping is
        done during the centralized training phase, we can assume we
        have access to the entire augmented rollout namely
        $\tilde{s}_t=(\bar{s}_t,\bar{q}_t,\bar{v}_t)$ at any given
        $t\in [0,T]$.  From the monitor reward function $\rho$, we can
        determine the weighting for a complete augmented rollout as $$
        \tilde{R}^i(\tilde{\zeta}_m) =
	\begin{cases}
	\rho(\bar{s}_T,q^i_T,v^i_T), & \text{if}\ q^i_T \in F \\
	-\infty & \text{otherwise}
	\end{cases}
	$$
	
	
	\begin{theorem}
		(Proof in Appendix Sec. \ref{app:proofs}.)
		\label{mkvgspec-thm}
		For any Markov game $\mathcal{M}_g$, specification
                $\phi$ and rollout $\zeta_m$ of $\mathcal{M}_g$,
                $\zeta_m$ satisfies $\phi$ if and only if there exists
                an augmented rollout $\tilde{\zeta}_m$ such that i)
                $\tilde{R}^i(\tilde{\zeta}_m) >0~\forall~i\in
                \mathcal{N}$ and ii) $\project(\tilde{\zeta}_m) =
                \zeta_m$.
	\end{theorem}
	
	The $\tilde{R}^i$ specified is $-\infty$ unless
        a trajectory reached a final state of the composite task
        monitor. To reduce the sparsity of this reward signal, we
        transform this into a shaped reward $\tilde{R}_s^i$ that gives
        partial credit to completing subtasks in the composite task
        monitor.
	
	Define for a non-final monitor state $q\in Q\setminus F$,
        function $\alpha:\mathcal{S}\times Q \times V \rightarrow
        \mathbb{R}$.
	$$\alpha(\bar{s},q,v) = \underset{(q,\sigma,u,q')\in \Delta,
          q\ne q' }{\max} \quantpred{\sigma}(\bar{s},v)$$
	
	\noindent This represents how close an augmented state
        $\tilde{s}=(\bar{s},q,v)$ is to transition to another state
        $\tilde{s}'$ with a different monitor state. Intuitively, the
        larger $\alpha$ is, the higher the chance of moving deeper
        into the task monitor. In order to use this definition on all
        $\sigma$, we overload $\sigma_l$ to also act on elements
        $\bar{s}=\{s^i\}_{i\in\mathcal{N}}\in\mathcal{S}$ by yielding
        for agent $i$, the value $\sigma_l(\bar{s}) = \sigma_l(s^i)$.
	
	Let $C_l$ be a lower bound on the final reward at a final
        monitor state, and $C_u$ being an upper bound on the absolute
        value of $\alpha$ over non-final monitor states.
	Also for $q\in Q$, let $d_q$ be length of the longest path from
        $q_0$ to $q$ in the graph $M_\phi$ (ignoring the self-loops in
        $\Delta$) and $D=\max_{q\in Q}d_q$. For an augmented rollout
        $\tilde{\zeta}_m$ let $\tilde{s}_k =
        (\bar{s}_k,q^i_k,\bar{v})$ be the first augmented state in
        $\tilde{\zeta_m}$ such that
        $q^i_k=q^i_{k+1}=\cdots=q^i_T$. Then we have the shaped
        reward,
	\begin{equation}
		\tilde{R}_s^i(\tilde{\zeta}_m) = 
		\begin{cases}
			 \max_{k\le j<T} \alpha(\bar{s}_j,q^i_T,v_j)  
				 + 2C_u\cdot(d_{q^i_T}-D) + C_l
			& \text{if}\ q^i_T \notin F \\
			\tilde{R}^i(\tilde{\zeta}_m) & \text{otherwise}
		\end{cases}
		\label{eq:reward}
	\end{equation}

	\begin{theorem} (Proof in Appendix Sec. \ref{app:proofs}.)
		\label{thm:rewardshaping}
		For two augmented rollouts
                $\tilde{\zeta}_m,\tilde{\zeta}_m'$, \\(i) if
                $\tilde{R}^i(\tilde{\zeta}_m) >
                \tilde{R}^i(\tilde{\zeta}_m')$, then 
                $\tilde{R}^i_s(\tilde{\zeta}_m) >
                \tilde{R}^i_s(\tilde{\zeta}_m')$, and (ii) if
                $\tilde{\zeta}_m$ and $\tilde{\zeta}_m'$ end in
                distinct non-final monitor states $q^i_T$ and
                $(q^i_T)'$ such that $d_{q^i_T} > d_{(q^i_T)'}$, then
                $\tilde{R}^i_s(\tilde{\zeta}_m) \geq
                \tilde{R}^i_s(\tilde{\zeta}_m')$.
	\end{theorem}

\begin{figure}[h]
	\begin{center}
		\includegraphics[width=.5\linewidth]{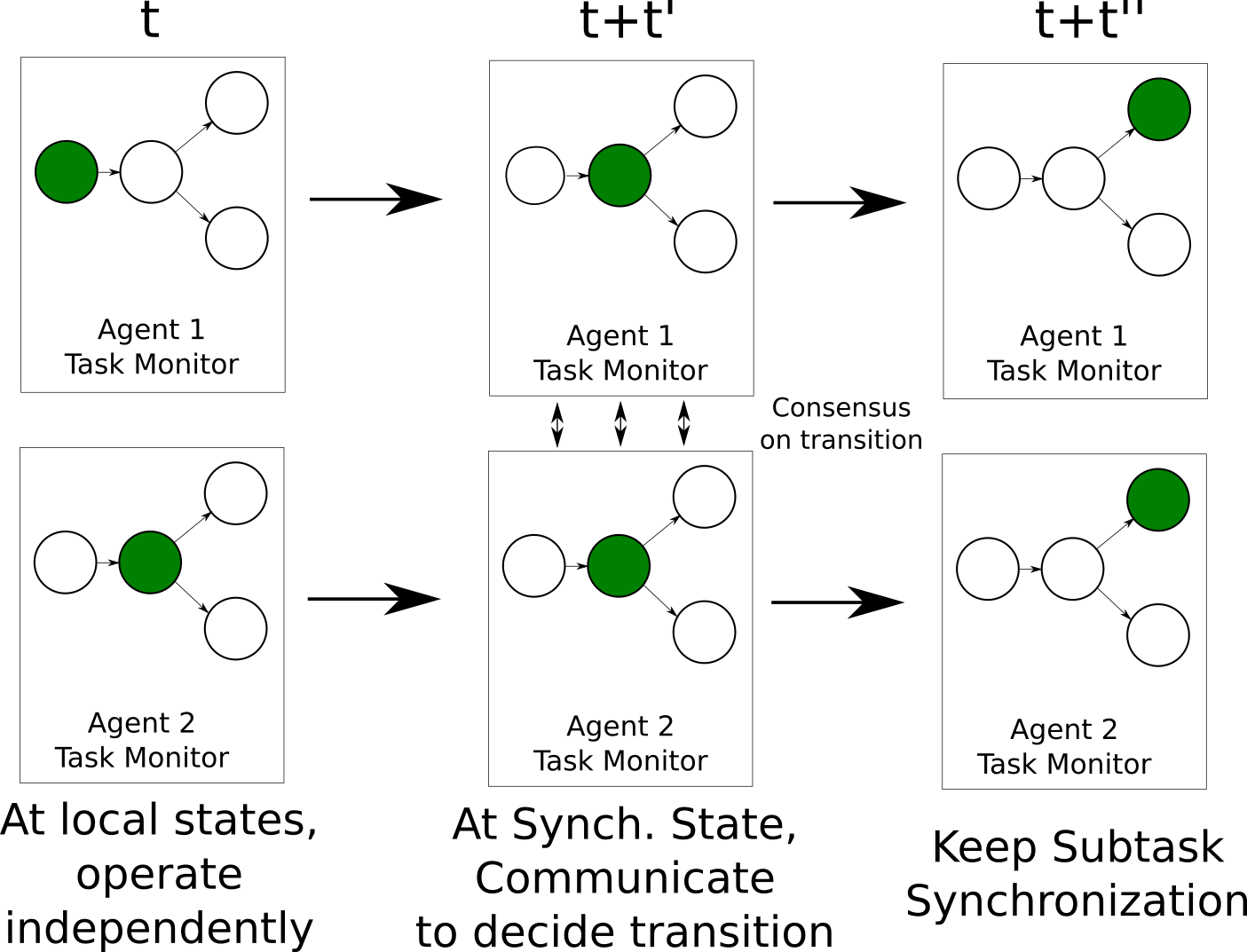}
	\end{center}
	\caption{Overview of the \toolname process for task synchronization. 
		Branching in the task monitor diagram denotes potential non-deterministic choices
		between future tasks (such as in $\mathbf{\phi_{ex}}$).
	Left to right represents the order of policy actions over a trajectory.
	Green states represent the current monitor state of that agent.}
	\label{fig:overview}
\end{figure}
	
	\section{Sub-task Synchronization}

	\subsubsection*{\textbf{Importance of Task Synchronization}}
	\label{ssec:task_synch}
	
	Consider the following example specification:
	$$ \phi_{1a}' =  \mathtt{reach}_{lo}(P) \  \mathtt{or}\  \mathtt{reach}_{gl}(Q) $$
	
	\noindent where $P,Q$ are some goals. To ensure flexibility
	with respect to the possible acceptable rollouts within
	$\phi_{1a}'$, the individual agent policies $\pi^i$ are
	learnable and the task transition chosen is dependent on the agent-specific observations. This flexibility between
	agents however, adds an additional possible failure method in
	achieving a global specification -  if even a single agent
	attempts to fulfill the global objective while the others
	decide to follow their local objectives, the specification
	would never be satisfied.

	\subsubsection{\textbf{Identifying Synchronization States}}
	\label{ssec:id_synch}

	As emphasized above, task
        synchronization is an important aspect of deploying these
        composite task monitors in the Markov game $\mathcal{M}_g$
        with specification $\phi$.  
        We show the existence of a subset
        of monitor states $\synchset\in Q$ where in order to maintain
        task synchronization, agents simply require a consensus on
        which monitor transition $\delta=(q,\sigma,u,q')$ to take. If
        we use $Q_g$ to symbolize the set of global monitor states, viz. all $q\in Q$ such that $\exists
        (q,\sigma_g,u_g,q')\in \Delta_g$, then we see that
        $Q_g\subseteq\synchset$. \edit{A valid choice for $q\in\synchset$ with $q\notin Q_g$ is all 
        	branching states in the graph of $M_\phi$ with a set refinement 
        	presented in the Appendix (Sec. \ref{app:ss_synch}).}
	
	During training, we enforce the condition that when an agent $i$ 
        has monitor state $q_t^i\in\synchset$, it must wait for time
        $t_1>t$ such that $q_{t_1}^j=q_t^i~\forall j \in \mathcal{N}$
        and then choose a common transition as the other agents. This
        is done during the centralized training phase by sharing the
        same transition between agents based on a majority vote.

	\section{Multi-Agent Specification Properties}
	\label{sec:ma_spec_prop}
	Consider a specification $\phi$ and let $  \mathcal{N} = \{1,\ldots,N\} $ be the set of all agents with $\zeta_m$ being a trajectory sampled from the environment. $\phi(\zeta_m, n)$ is used to denote that the specification is satisfied on $\zeta_m$ for the set of agents $n \subseteq \mathcal{N}$ (i.e. $\pred{\phi(\zeta_m, n)} == \true$).
	
	\subsubsection{\textbf{MA-Distributive}}
	Many specifications pertaining to MA problems can be satisfied independent of the number of agents. 
	At its \edit{core}, we have the condition that a specification being satisfied with respect to 
	a union of two disjoint sets of agents implies that it can be satisfied on both sets independently.
	  Namely \edit{if $n_1,n_2\subset\mathcal{N}$ with $n_1\cap n_2 = \emptyset$} then an MA-Distributive specification satisifies the following condition:
	
	$$\phi(\zeta_m,n_1\cup n_2) \implies \phi(\zeta_m,n_1) \land \phi(\zeta_m,n_2) $$
	
	\subsubsection{\textbf{MA-Decomposable}}
	Certain specifications satisfy a decomposibility property particular to 
	multi-agent problems that can help in scaling with respect to the number of agents.
	
	 Say $ \exists\ k \in \{1, \ldots,  N-1\}$ such that
	$$\phi(\zeta_m, \mathcal{N}) \implies \phi_k(\zeta_m, \mathcal{N}) = \bigwedge\limits_{j\in \{1,\ldots,J\}} \phi(\zeta_m,n_j) $$
	where
	$$ n_j \subset \mathcal{N} \ , \ k \leq |n_j| < N \ ,\ J = \lfloor\frac{N}{k}\rfloor\ ,
	\ \bigcap\limits_j n_j = \emptyset \ ,\  \bigcup\limits_j n_j = \mathcal{N} $$
	with $\lfloor\rfloor$ representing the floor function. Each $n_j$ is a set of at least $k$ unique agents and $\{n_j\}_j$ forms a partition over $\mathcal{N}$.
	
	We then call the specification $\phi$ \emph{MA-Decomposable} with decomposibility factor $k$. 
	Here $\phi_k$ can be thought of as a means to approximate the specification $\phi$ to smaller
	 groups of agents within the set of agents $\mathcal{N}$. 
	Provided we find a value of $k$, we can then use this as the basis of our
	 \scalingname scaling method to significantly improve training times for larger numbers of agents.
	
	\begin{theorem} (Proof in Appendix Sec. \ref{app:proofs}.)
		\label{thm:distributivedecomp}
		All MA-Distributive specifications are also MA-Decomposable with decomposability factors $k\in\mathbb{Z}^+, 1\le k <N$. 
	\end{theorem}
	Notably all compositions of $\mathtt{reach}_{gl}$ and  $\mathtt{reach}_{lo}$
	 within our language are MA-Distributive and are thus MA-Decomposable with factor $k=2$.
	  \footnote{ It is satisfied with $k=1$ as well but this is the trivial case where $\mathtt{reach}_{gl}$ and  $\mathtt{reach}_{lo}$ are equivalent. }
	This is far from a general property however, as one can define specifications on $N$ robots
	such as $\mathtt{achieve}$("collect $x$ fruits") where each robot can carry 
	at most $x/N$ fruits . In this case, no single subset of agents 
	can satisfy the specification as the total capacity of fruits would be less 
	than $x$ and the specification is neither MA-Distributive nor MA-Decomposable.

	\section{Algorithm}
	\subsubsection*{\textbf{Training}}
	
	Agents learn $\pi^i(s^i,v^i,q^i)=(a^i,\delta^i)$ on the
        augmented Markov game $\mathcal{M}_g'$ where $s^i,v^i,q^i$ are
        agent-specific state, register value and task monitor state
        respectively.  Since training is centralized, all agent
        task monitors receive the same global state.  Based on our
        discussion in Sec.~\ref{ssec:task_synch}, if an agent is in any
        given global monitor state, we wait for other agents to enter the same
        state, then do the $\arg \max$ task transition for all agents
        in the same state. In addition, at the synchronization states
        (Sec. \ref{ssec:id_synch}), we perform a similar process to
        select the task transition.  These trained \textit{augmented
          policies} are then projected into policies that can act in
        the original $\mathcal{M}_g$.
        
    \begin{figure}[!h]
    	\begin{center}
    		\includegraphics[width=0.4\linewidth]{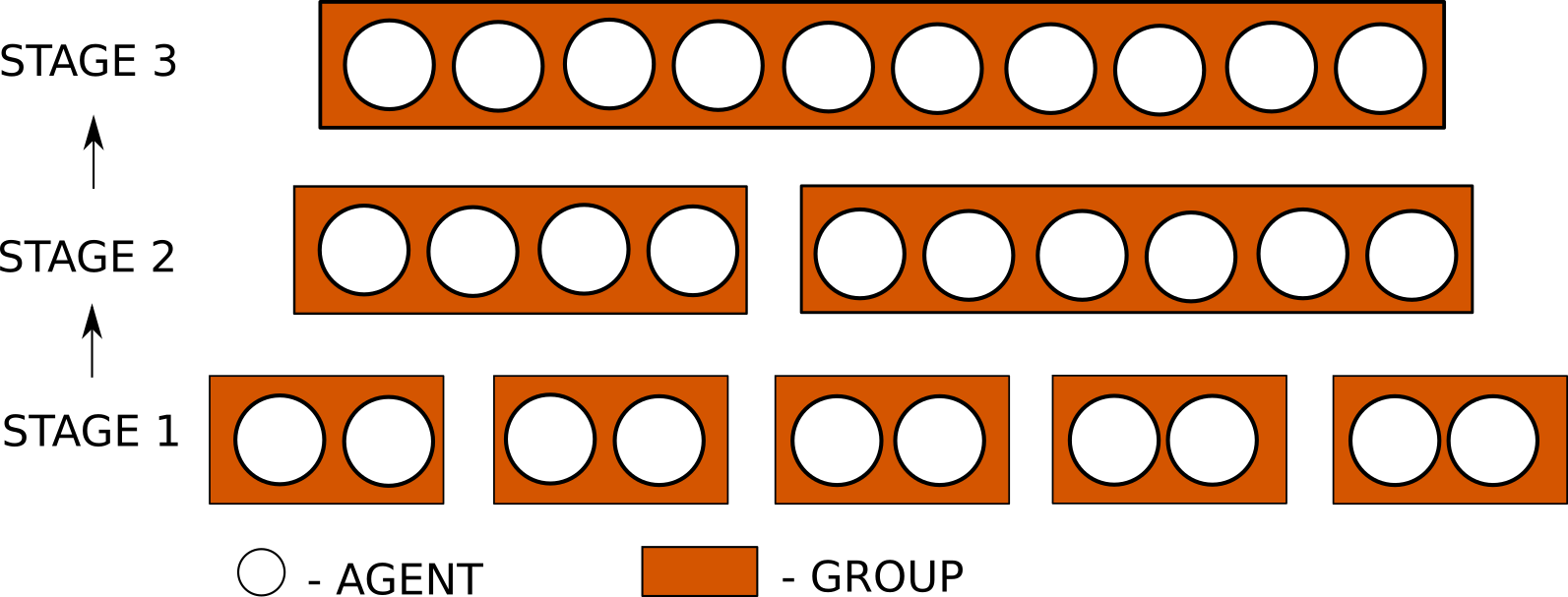}
    	\end{center}
    	\caption{Example \scalingname Scaling Process with $N=10,k=2,f=2$ on an MA-Decomposable spec $\phi$ with decomposability factor $2$. At Stage 1, $g_1=2$ to start with 5 groups. Next $g_2=fg_1 = 4$ which forms 2 groups. Finally at Stage 3, $g_3 = fg_2=8$ which forms one group ($\mathcal{N}$).}
    	\label{fig:example_scaling}
    \end{figure}    
 
 \subsubsection*{\textbf{Scaling MA-Decomposable Specifications}}
 \label{sec:scaling_method}
 Our algorithm for scaling based off the MA-Decomposable property is shown in Alg. 
 \ref{algo:scaling} \edit{(refer Appendix)} and we name it \scalingname scaling.
 Essentially, we approximate the spec. $\phi$ by first independently considering smaller groups within the larger set of agents $\mathcal{N}$ and try to obtain a policy satisfying $\phi$ on these smaller groups.
 By progressively making the group sizes larger over stages and repeating the policy training
  process while continuing from the previous training stage's policy parameters , 
  we form a curriculum that eases solving the original problem $\phi$ on all agents $\mathcal{N}$.
 
 In Fig. \ref{fig:example_scaling} we demonstrate \scalingname scaling for $N=10$ agents on a spec. $\phi$ which is MA-Decomposable
 with decomposability factor $2$. For this example we set the scaling parameters $k=2$ and $f=2$.
Initially we have a min. group size $g_1 = 2$ and this is changed to
$g_2 = 4$ and $g_3 = 8$ from setting the scaling factor.  We increment
the stage number every time all the groups of a stage have satisfied
the entire specification $\phi$ w.r.t. their group.  While separating
training into stages, agents must be encouraged to move from stage $i$
to stage $i+1$. To ensure this, we need to scale rewards based on the
stage. We chose a simple linear scaling where for stage number $i$ and
time step $t$, each agent receives reward $ r_{i,t} = ic_k + CTM_{i,t}$ where 
$CTM_{i,t}$ is the original composite task monitor reward at stage $i$
and $c_k\in\mathbb{R}$ is a constant.  By bounding the reward terms such
that rewards across stages are monotonically increasing  $(r_{i,t}<r_{i+1,t'})$ we can find a suitable $c_k$ to be $(2D+1)C_u$
(refer Appendix Sec. \ref{app:scaling_const}) where the terms are the same as in
Eq. \ref{eq:reward}.
 
 From setting the initial min. group size $g_1$ and scaling factor
 $f$, we get the total number of learning stages ($T_s$) as $T_s =
 \lfloor \log_f(N) - \log_f(k) \rfloor = \mathcal{O}(\log_f(N))$.  We
 build the intuition behind why \scalingname scaling is effective in
 the Appendix (Sec. \ref{app:scaling_int}), by describing it as a form
 of curriculum learning.
 
	\subsubsection*{\textbf{Deployment}}
	Policies are constructed to proceed with only local
        information ($s^i,v^i,q^i$). Since we cannot share the whole
        system state with the agent policies during deployment yet our
        composite task monitor requires access to this state at all
        times, we allow the following relaxations: 1) Global
        predicates $\sigma_g(\bar{s},v)$ enabling task monitor
        transitions need global state and access it during deployment.
        2) Global register updates $u_g(\bar{s},v)$ are also a
        function of global state and access it during deployment.

In order to maintain task synchronization, agents use a consensus
based communication method to decide task monitor transitions at
global and synchronization states. If agents choose different task transitions
at these monitor states, the majority vote is used as done during training.

	\section{Experiment Setup}
	
	Our experiments aim to \edit{ validate that the use }of a distributed
        task monitor can achieve synchronization during the
        deployment of multiple agents on a range of specifications.
	
	In addition, to emphasize the need for distribution of task
        monitors to alleviate the state space explosion caused by
        mixing local and global specifications, we include experiments
        with \spectrl applied to a centralized controller.
	
	Lastly, we provide results showing the efficacy of the
        \scalingname scaling approach for larger numbers of agents
        when presented with a specification that satisfies the
        MA-Decomposability property (Sec. \ref{sec:ma_spec_prop}).
	
	As a baseline comparison, we also choose to run our algorithm
        without giving policies access to the monitor state
        (\textbf{no\_mon}). These are trained with the same shaped
        reward as \toolname.
        We also provide a Reward Machine baseline (\textbf{RM}) for $\phi_1$ with continuous rewards
        since $\phi_1$ is similar to the 'Rendezvous' specification in \cite{10.5555/3463952.3464063}.

	\subsubsection{\textbf{Environment}} Our first set of experiments are done on a 2D Navigation
        problem with $N=3$ agents.  The observations ($\mathcal{S}\in
        \mathbb{R}^2$) used are coordinates within the space 
        with the action space ($\mathcal{A}\in
        \mathbb{R}^2$) providing the velocity of the agent.

	The second set of experiments towards
        higher dimension 3D benchmarks, represent particle motion in a 3D
        space. We train multiple agents ($N=3$) in the 3D space
        ($\mathcal{S}\in \mathbb{R}^3$) with a 3D action space
        ($\mathcal{A}\in \mathbb{R}^3$) to show the scaling potential
        of our framework.
	
	\edit{The final set of experiments were on a modern discrete-action MARL benchmark built in Starcraft 2 \cite{smac_aamas} 
 with $N=8$ agents (the "8m" map). Each agent has 14 discrete actions with a state space $\mathcal{S}\in \mathbb{R}^{80}$ 
	representing a partial view of allies and enemies.}

	\subsubsection{\textbf{Algorithm Choices}}  
\begin{figure}[!h]
				\begin{center}
					\minipage{0.4\linewidth}
					\includegraphics[width=\linewidth]{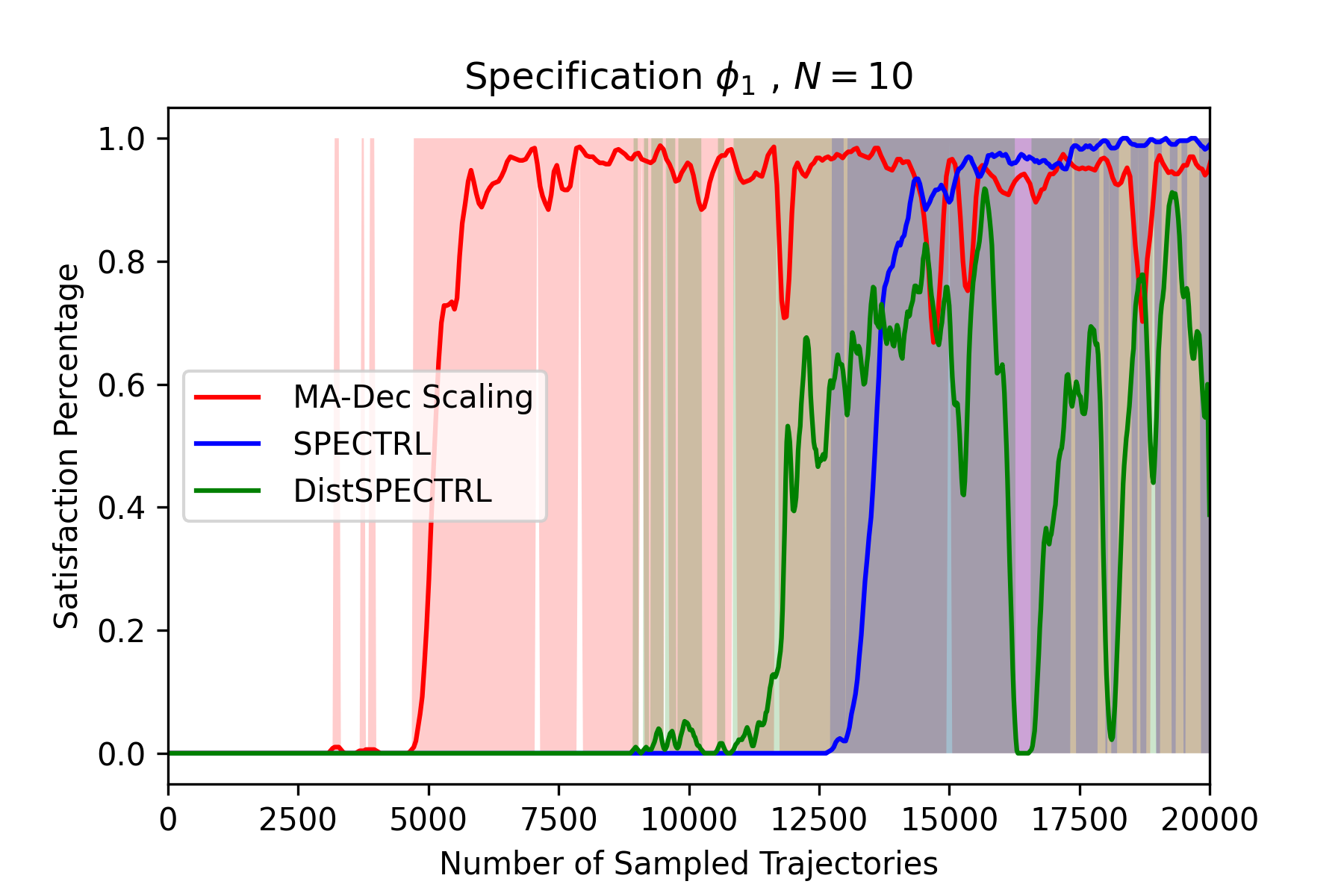}
					\endminipage
					\minipage{0.4\linewidth}
					\includegraphics[width=\linewidth]{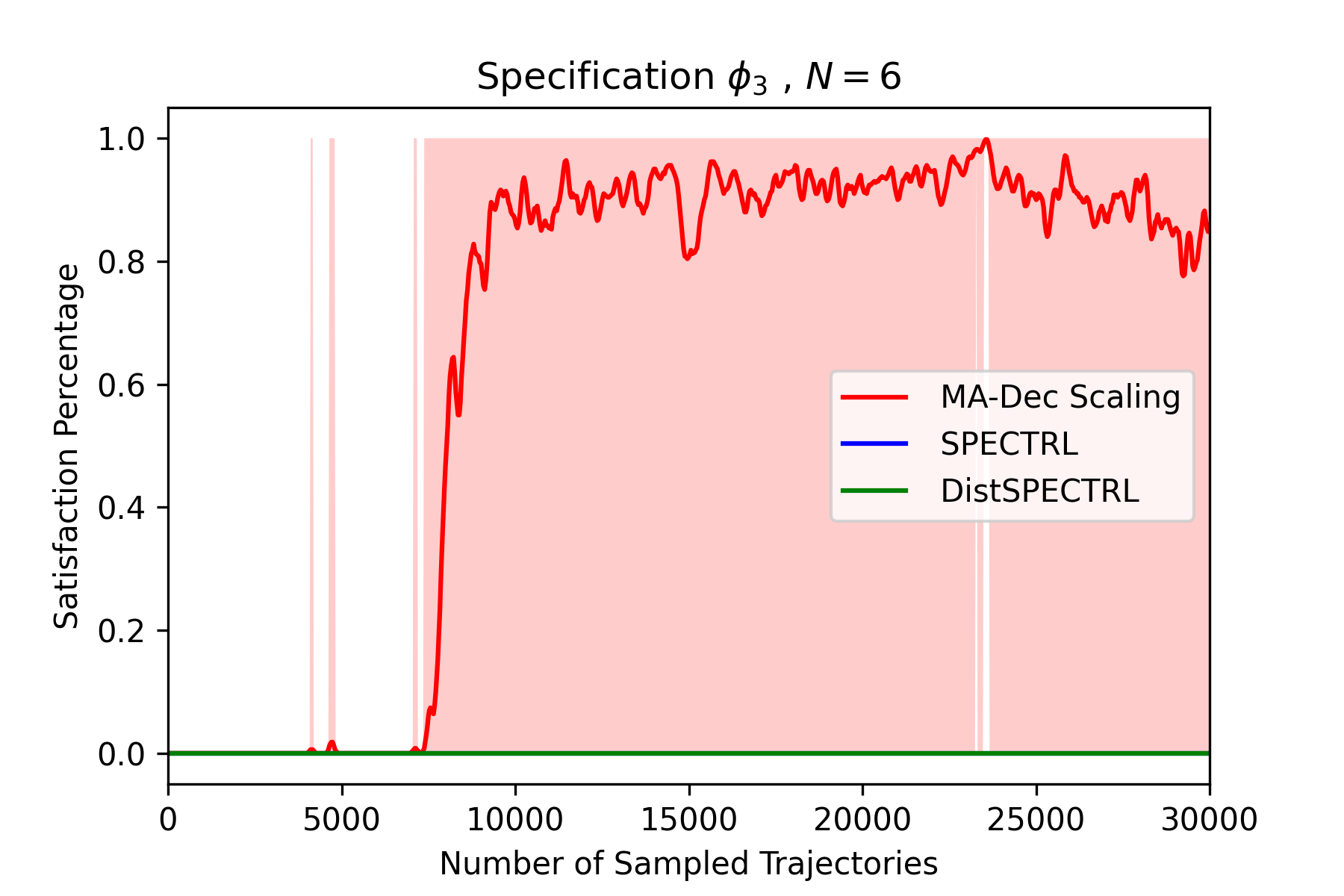}
					\endminipage
					\captionof{figure}{Specification satisfaction percentages for $N=10$ agents on $\phi_1$ (left) and $N=6$ agents on $\phi_3$ (right) comparing the \scalingname scaling (red) to centralized \spectrl (blue) and vanilla \toolname (green) i.e. without scaling enhancements.}
					\label{fig:scaling_results}
				\end{center}
			\end{figure}

For the scaling experiments (Fig. \ref{fig:scaling_results})
        we used the 2D Navigation problem with horizon $T=500$ and the
        scaling parameters\footnote{While we could start with $k=1$,
          we set $k=2$ to reduce the number of learning stages.} $k=2$
        and $f=2$. We also choose a version of PPO with a centralized
        Critic to train the augmented Markov Game noting that our
        framework is agnostic to the choice of training algorithm.
        The current stage is passed to the agents as an extra integer
        dimension.  
        For other experiments we chose PPO with independent critics as
        our learning algorithm.  Experiments were implemented using
        the RLLib toolkit \cite{pmlr-v80-liang18b}.
	
	\subsubsection*{\textbf{Specifications}}
	
	 \textbf{(2D Navigation)} The evaluated specifications are a mix of local and global objectives. The reach predicates have an error tolerance of 1 (the $L_\infty$ distance from the goal).
	
	(i) $ \phi_1 = \mathtt{reach_{gl}}(5,0); \mathtt{reach_{gl}}(0,0)$ 
	, (ii) $ \phi_2 = \phi_1; \mathtt{reach_{gl}}(3,0)$ 
	
	(iii) $ \phi_3 = \mathtt{reach_{lo}}(5,0);\mathtt{reach_{gl}}(0,0); \mathtt{reach_{gl}}(3,0)$ 
	
	(iv) $ \phi_4 = \left[\mathtt{reach_{lo}}(3,0) \ \mathtt{or}\ \mathtt{reach_{lo}}(5,10) \right] ; \phi_3$ 

\noindent\edit{
\textbf{(SC2)}  
$\phi_{sc}$ represents 'kiting' behaviour and is explained further in the Appendix (Sec. \ref{app:sc2}).
 $ \phi_{sc} = \phi_{sc_a};\phi_{sc_a};\phi_{sc_a} $ where $
	\phi_{sc_a}=  \mathtt{away\_from\_enemy}_{gl};  \mathtt{shooting\_range}_{lo};
$}

	\noindent\textbf{(3D Environment )} 
	$ \phi_a = \mathtt{reach_{lo}}(5,0,0);\mathtt{reach_{gl}}(0,0,0); \mathtt{reach_{gl}}(3,0,0)$ 
	\edit{is the specification considered within X-Y-Z coordinates.}

		\begin{figure*}[!h]
		\includegraphics[width=\linewidth]{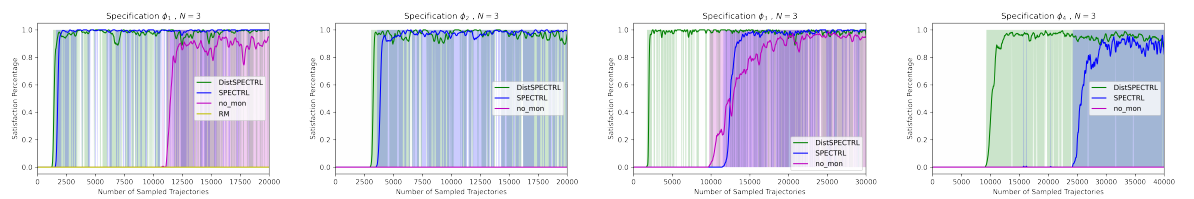}
		\caption[Caption for LOF]{Satisfaction percentages on specifications  $\phi_1$,$\phi_2$,$\phi_3$ and $\phi_4$ with 
			$N=3$ agents. The shaded regions show the maximum and minimum achieved over 5 separate evaluation runs }
		\label{fig:results}
	\end{figure*}
	
			
			\begin{figure}[!h]
				\begin{center}
					\minipage{0.4\linewidth}
					\includegraphics[width=\linewidth]{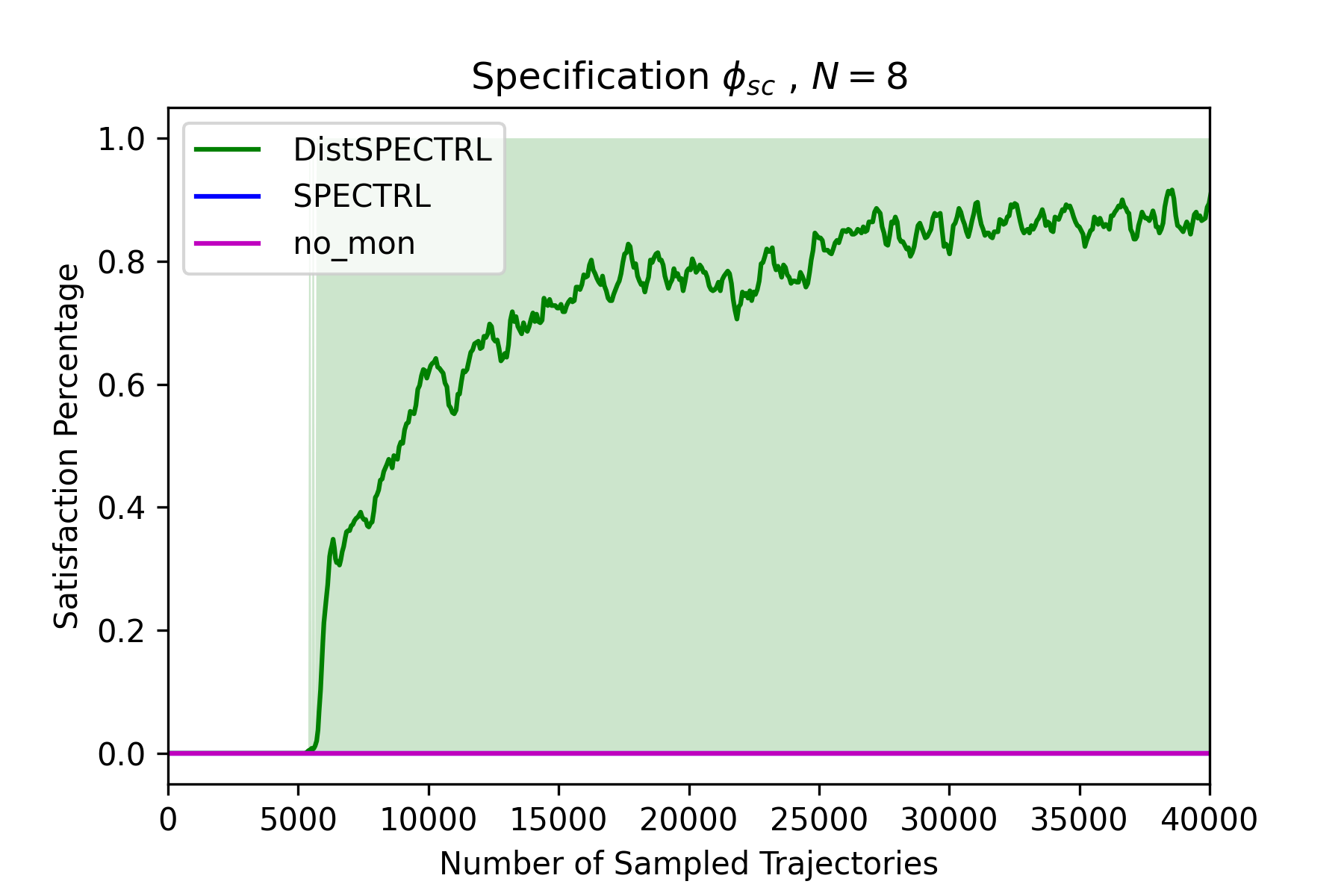}
					\endminipage
					\minipage{0.4\linewidth}
					\includegraphics[width=\linewidth]{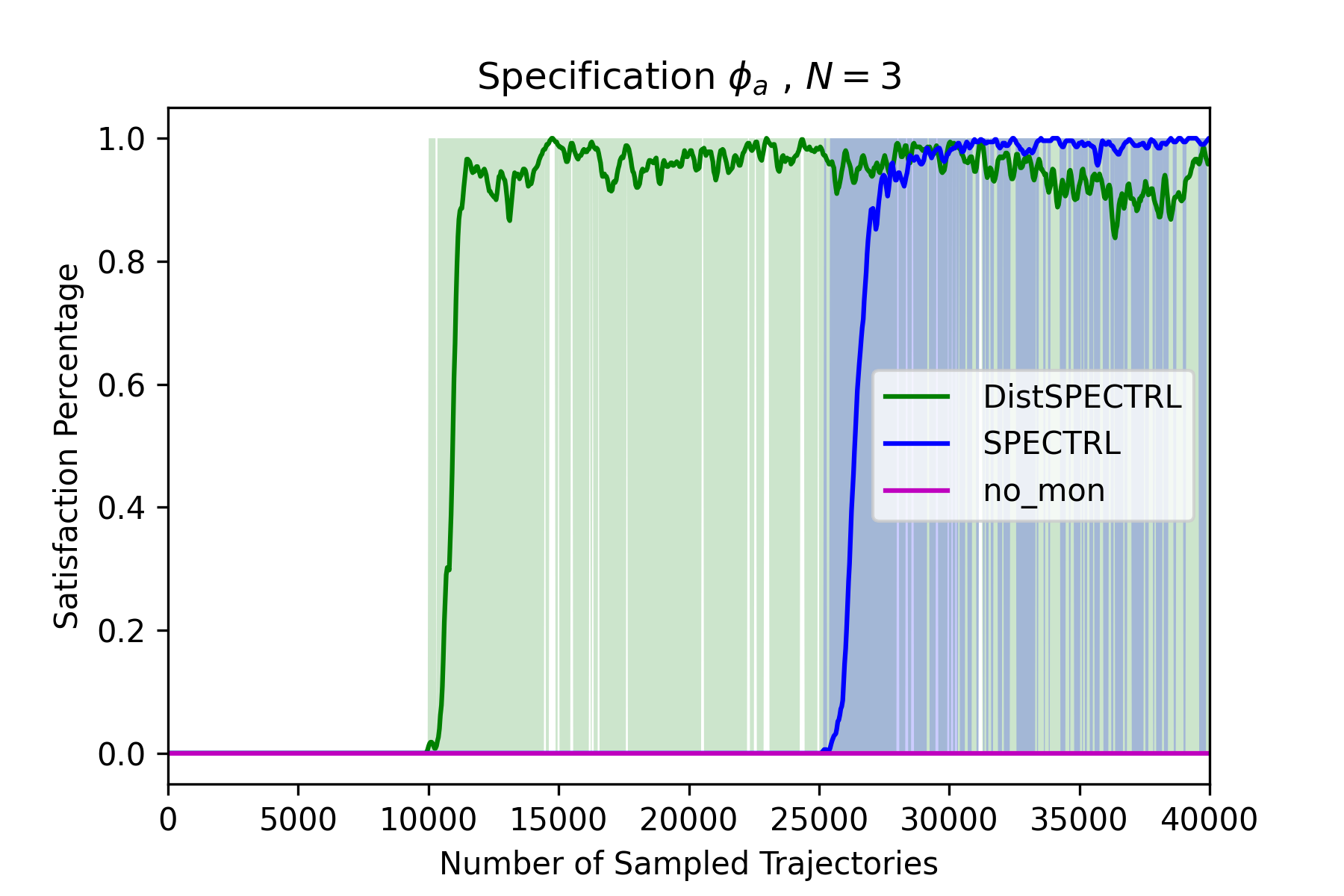}
					\endminipage
					\captionof{figure}{ \edit{Specification satisfaction percentages (left) for the StarCraft 2 specification $\phi_{sc}$ with $N=8$ agents and (right) for the 3D Navigation experiments on specification  $\phi_a$. }}
					\label{fig:ant_results}
				\end{center}
			\end{figure}
			
	\begin{table}[h]
		\centering
		\minipage{0.44\linewidth}
		\captionof{table}{Specification satisfaction percentages on convergence for Fig.\ref{fig:results},\ref{fig:ant_results}}
		\begin{tabular}{c|ccc}
			\toprule
			Spec.  & \toolname & no\_mon & \spectrl \\
			\midrule
			$\phi_{1}$       & 99.62  & 91.17 & 100.00   \\
			$\phi_{2}$  & 99.05 &  00.00    & 97.38   \\
			$\phi_{3}$     &  97.59 &  94.77 & 96.81  \\
			$\phi_{4}$     &  97.31 &   00.00 & 90.78  \\
			\midrule
			$\phi_{a}$     &  98.49 &   00.00 & 99.60   \\
			\midrule
			$\phi_{sc}$    & 86.79 & 00.00 & 00.00 \\
			\bottomrule
		\end{tabular}
		\label{tab:results_table}
		\endminipage
		\hfill
		\minipage{0.55\linewidth}
		\captionof{table}{Specification satisfaction percentages on convergence for Fig.\ref{fig:scaling_results}, (Scaling to more Agents)}
		\begin{tabular}{c|ccc}
			\toprule
			Spec. $/$ \# Agents  & MA-Dec & \toolname & \spectrl \\
			\midrule
			$\phi_{3}/ N$ =6     & 94.09  & 0.00 & 0.00    \\
			\midrule
			$\phi_{1}/ N$ =6    & 97.83  & 80.67 & 98.96    \\
			$\phi_{1}/ N$ =10      & 97.03  & 72.30 & 99.28  \\
			\bottomrule
		\end{tabular}
		\label{tab:scaling_results}
		\endminipage

	\end{table}
	\section{Results}
	\subsubsection{\textbf{Handling Expressive Specifications}}
	The experiments in Fig. \ref{fig:results} demonstrate
        execution when the task monitor predicates have access to the
        the entire system state. This provides agents with information
        sufficient to calculate global predicates for task
        monitor transitions.
	\edit{The overall satisfaction percentage is reported with the value
        $0$ being an incomplete task to $1.0$ being the entire
        specification satisfied. }
	
	While \spectrl has often been shown to be more effective
        \cite{NIPS2019_9462,DBLP:journals/corr/abs-2106-13906} than many existing methods (e.g. \textbf{RM} case) for task
        specification, the further utility of the monitor state in
        enhancing coordination between agents is clearly evident in a
        distributed setting. 
	\edit{The task monitor state is essential for
        coordination as our baseline \textbf{no\_mon} is often unable to complete the entire task
         (even by exhaustively going through possible
        transitions) and global task completion requires enhanced levels
        of synchronization between agents.}

	From Table \ref{tab:results_table} we see that upon
        convergence of the learning algorithm, the agent is able to
        maintain a nearly 100\% task completion rate for our tested
        specifications, a significant improvement in comparison to the
        \textbf{no\_mon} case, showing the importance of the task
        monitor as part of a multi-agent policy.
	
	\subsubsection{\textbf{Benefits of Distribution Over Centralization}}
	\edit{The centralized \spectrl graphs (blue curves in
        Figs. \ref{fig:results}, \ref{fig:ant_results}, \ref{fig:scaling_results}) show that while distribution may not
        be necessary for certain specifications with few local
        portions (e.g. $\phi_2$), concatenating them will
        quickly lead to learning difficulties with larger 
        number of agents (Fig. \ref{fig:ant_results}, $\phi_{sc}$ and Fig. \ref{fig:scaling_results}, $\phi_3$).
		} This difficulty is due
        in large part to state space explosion of the task monitor in
        these cases as is apparent by the significantly better
        performance of our distributed algorithm.
        \edit{We also remind the reader that a centralized algorithm is further
        	disadvantageous in MARL settings due to the added synchronization cost between agents
        	during deployment.}


\subsubsection{\textbf{Scaling to Larger State Spaces}}
	The results in Fig. \ref{fig:ant_results} show promise that
        the \toolname framework can be scaled up to larger dimension
        tasks as well. The 3D environment results exhibits similar
        behavior to the 2D case with the \textbf{no\_mon} showing 
        difficulty in progressing beyond the local tasks
	in the larger state space with sparser predicates. 
	\edit{
	The $\phi_{sc}$ results also show promise in defining relevant predicates and achieving general 
	specifications for modern MARL benchmarks.}
	
	\subsubsection{\textbf{Scaling to More Agents}}
	Table \ref{tab:scaling_results} and Fig. \ref{fig:scaling_results} demonstrate the benefits of \scalingname 
	scaling for larger $N$ when presented with an MA-Decomposable specification.
	At smaller ranges of $N$ as well as less complex combinations of mixed and global objectives, 
	the effect of \scalingname scaling is less pronounced. 
	We observe that the stage based learning is crucial for a even simple mixed specification like $\phi_3$ with as little
	as $N=6$ agents.

	%
	
	\section{Related Work}
	

	Multi-agent imitation learning
        \cite{NEURIPS2018_240c945b,pmlr-v70-le17a,pmlr-v97-yu19e} uses
        demonstrations of a task to specify desired behavior. However
        in many cases, directly being able to encode a specification
        by means of our framework is more straightforward and removes
        the need to have demonstrations beforehand. Given
        demonstrations, one may be able to infer the specification
        \cite{NEURIPS2018_74934548} and make refinements or
        compositions for use in our framework.

	
	TLTL\cite{8206234} is another scheme to incorporate temporal
        logic constraints into learning enabled controllers, although
        its insufficiency in handling non-Markovian specifications led
        us to choose \spectrl as the basis for our methodology.
	Reward Machines (RMs)
        \cite{tor-etal-arxiv20,cam-etal-ijcai19,tor-etal-icml18} are
        an automaton-based framework to encode different tasks into an
        MDP. While RMs can handle many non-Markovian reward
        structures, a major difference is that \spectrl starts with a
        logical temporal logic specification and includes with the
        automaton the presence of
	  memory (in the form of registers 
	 capable of storing real-valued information).
	 Recent work \cite{DBLP:journals/corr/abs-2106-13906} shows the 
	 relative advantages \spectrl -based solutions may have over a range of 
	 continuous benchmarks. 
	 
	 Concurrent work has introduced the benefits of a temporal
         logic based approach to reward specification
         \cite{10.5555/3463952.3464024}. While experimental results
         are not yet displayed, the convergence guarantees of the
         given algorithm are promising. Since we use complex
         non-linear function approximators (neural networks) in our
         work, such guarantees are harder to provide.  Reward Machines
         have also been explored as a means of specifying behavior in
         multi-agent systems \cite{10.5555/3463952.3464063} albeit in
         discrete state-action systems that lend themselves to
         applying tabular RL methods such as Q-learning.  One may
         extend this framework to continuous systems by means of
         function approximation but to the best of our knowledge, this
         has not been attempted yet.  Similar to our synchronization
         state, the authors use a defined local event set to sync
         tasks between multiple agents and requires being aware of
         shared events visible to the other agents.
	
  \edit{In the same spirit as our stage-based approach,} transferring learning from smaller groups of agents to larger
 ones has also been explored \cite{Wang_Yang_Liu_Hao_Hao_Hu_Chen_Fan_Gao_2020}.
 Lastly, while we chose PPO to train the individual agents \edit{for its simplicity}, our framework is
 agnostic to the RL algorithm used and can be made to work with other
 modern multi-agent RL setups
 \cite{10.5555/3295222.3295385,foerster2018a} for
 greater coordination capabilities.

	
	\section{Conclusion}
	We have introduced a new specification language to help detail
        MARL tasks and describe how it can be used to compile a
        desired description of a distributed execution in order to
        achieve specified objectives.  Our framework makes task
        synchronization realizable among agents through the use of: 1)
        Global predicates providing checks for task completion that
        are easily computed, well-defined and tractable; 2) A monitor
        state to keep track of task completion; and 3) Synchronization
        states to prevent objectives from diverging among agents.
		
	\paragraph{\textbf{Acknowledgements}}
	This work was supported in part by C-BRIC, one of six centers in JUMP, 
	a Semiconductor Research Corporation (SRC) program sponsored by DARPA.
	
	

	
	
	
	\bibliographystyle{splncs04}

	
	\include{sections/appendix}
	
\end{document}

%% file: sections/macros.tex
\newcommand{\BibTeX}{\rm B\kern-.05em{\sc i\kern-.025em b}\kern-.08em\TeX}

\newcommand\todo[1]{\textcolor{red}{[\texttt{TODO: #1}]}}
\newcommand\sj[1]{\textcolor{green}{[\texttt{SJ: #1}]}}

\newcommand\edit[1]{{#1}}

\newcommand\statemap{ \mathtt{MAP}}
\newcommand\pred[1]{ \llbracket#1\rrbracket}
\newcommand\quantpred[1]{ \llbracket#1\rrbracket_q}
\newcommand\approxpred[1]{ \llbracket#1\rrbracket_a}
\newcommand\synchset{ \mathtt{Sync}}
\newcommand\project{ \mathtt{proj}}
\newcommand{\neighbor}[1]{\mathcal{N}_g(#1)}
\newcommand{\true}{\mathtt{True}}
\newcommand{\false}{\texttt{false}}

\newcommand{\toolname}{Dist\textsc{Spectrl}\xspace}
\newcommand{\spectrl}{\textsc{Spectrl}\xspace}
\newcommand{\scalingname}{MA-Dec\xspace}

%% file: sections/appendix.tex
\appendix
\section{Global Task Monitor Construction}
\label{app:construction}
We refer readers to the local specification compilation rules of \spectrl (defined in their Appendix).
We highlight the main differences with mixed objectives here.

For a specification $\phi$ let $Q_g$ the set of states with global predicates in $\phi$. Note that $Q_g\subseteq \synchset$ as mentioned in Sec \ref{ssec:id_synch}. Here we consider a $\phi$ to be global if it contains any global predicates.
\subsection{$\mathtt{achieve}~b$ when $b$ is a global predicate}
Add both states to global states $Q_g$.

\subsection{ $\phi_1 ;~ \phi_2$ when $\phi_2$ is global}
If $\phi_2$ has an initial global state or $(q_0)_{\phi_2}\in (\synchset)_{\phi_2}$ then the transition from the final state of $\phi_1$ to $\phi_2$ is also global. If $q_a\in F_{\phi_1}$ then $q_a\in Q_g$.
\subsection{ $\phi_1 ;~ \phi_2$ when $\phi_1$ is global}
It is the same as the local case with $Q_g = (Q_g)_{\phi_1}$
\subsection{ $\phi_1 ;~ \phi_2$ when both $\phi_1,\phi_2$ are global}
It is the same as the local case with $Q_g = (Q_g)_{\phi_1}\cup (Q_g)_{\phi_2}$

\subsection{ $\phi_1 \mathtt{or}~ \phi_2$ when $\phi_2$ is global}
Without loss of generality, if $\phi_2$ contains global states then the common start state (as part of the compilation rules of $\mathtt{or}$) is a synchronization state. $Q_g = (Q_g)_{\phi_1}\cup (Q_g)_{\phi_2}$.

\section{Scaling MA Specifications}
\label{app:scaling_int}
\scalingname Scaling can be thought of as a form of curriculum learning for MA-Distributive specifications.
 We progressively narrow down the valid space of parameters that satisfy the specification
  $\phi_k$ by increasing the value of $k$ by a positive integer factor $f>1$.
 Consider a set of $N=10$ agents and an MA-Distributive specification $\phi$. $\phi$ is also MA-Decomposable with factor $8$ by Thm. \ref{thm:distributivedecomp}.
 Since the spec. is MA-Distributive as well $\phi_8(\eta,\mathcal{N}) \implies \phi_4(\eta,\mathcal{N}) \implies \phi_2(\eta,\mathcal{N})$.

\begin{figure}[!h]
	\begin{center}
		\includegraphics[width=.5\linewidth]{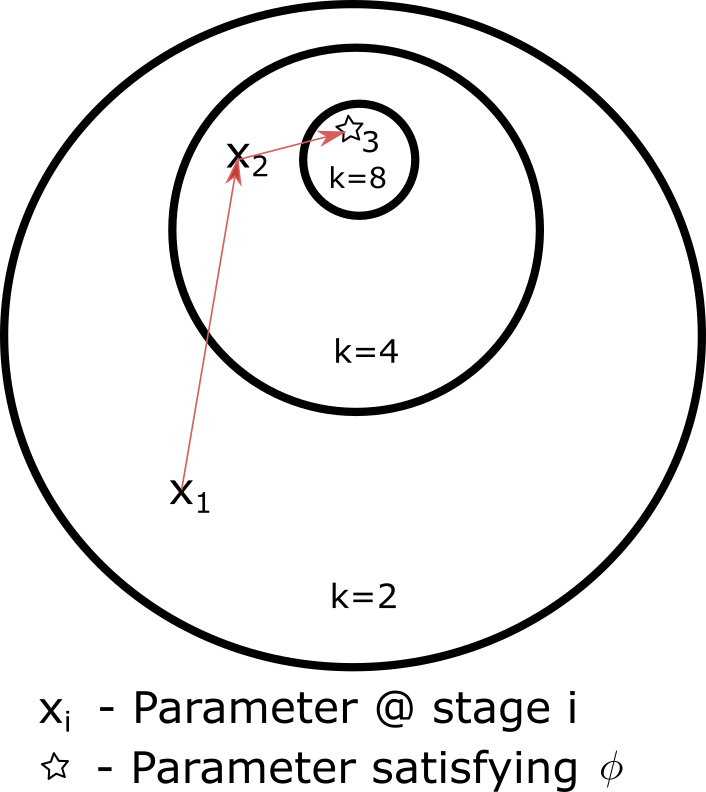}
	\end{center}
	\caption{Scaling intuition for $N=10,k=2,f=2$. We represent the policy parameter space and the respective placement of parameters that satisfy $\phi_k$ for various values of $k$. The arrows show the direction we proceed searching for parameters in our scaling process.}
	\label{fig:scaling_intuition}
\end{figure}

Intuitively, as shown in Fig. \ref{fig:scaling_intuition}, the policy parameter $\Pi_\theta$ satisfying $\phi_8$ will also satisfy $\phi_4$ and $\phi_2$ as groups of 8 agents can either be considered two groups of 4 agents or four groups of 2 agents.

Thus we position the parameter spaces as shown, and in the first stage attempt to find a parameter within the 
largest region satisfying $\phi_2(\mathcal{N})$.
As the learning progresses, the curriculum narrows down the desired search space until we obtain the 
parameters satisfying the specification $\phi(\mathcal{N})$.

\begin{figure}[!h]
	\begin{center}
		\includegraphics[width=.7\linewidth]{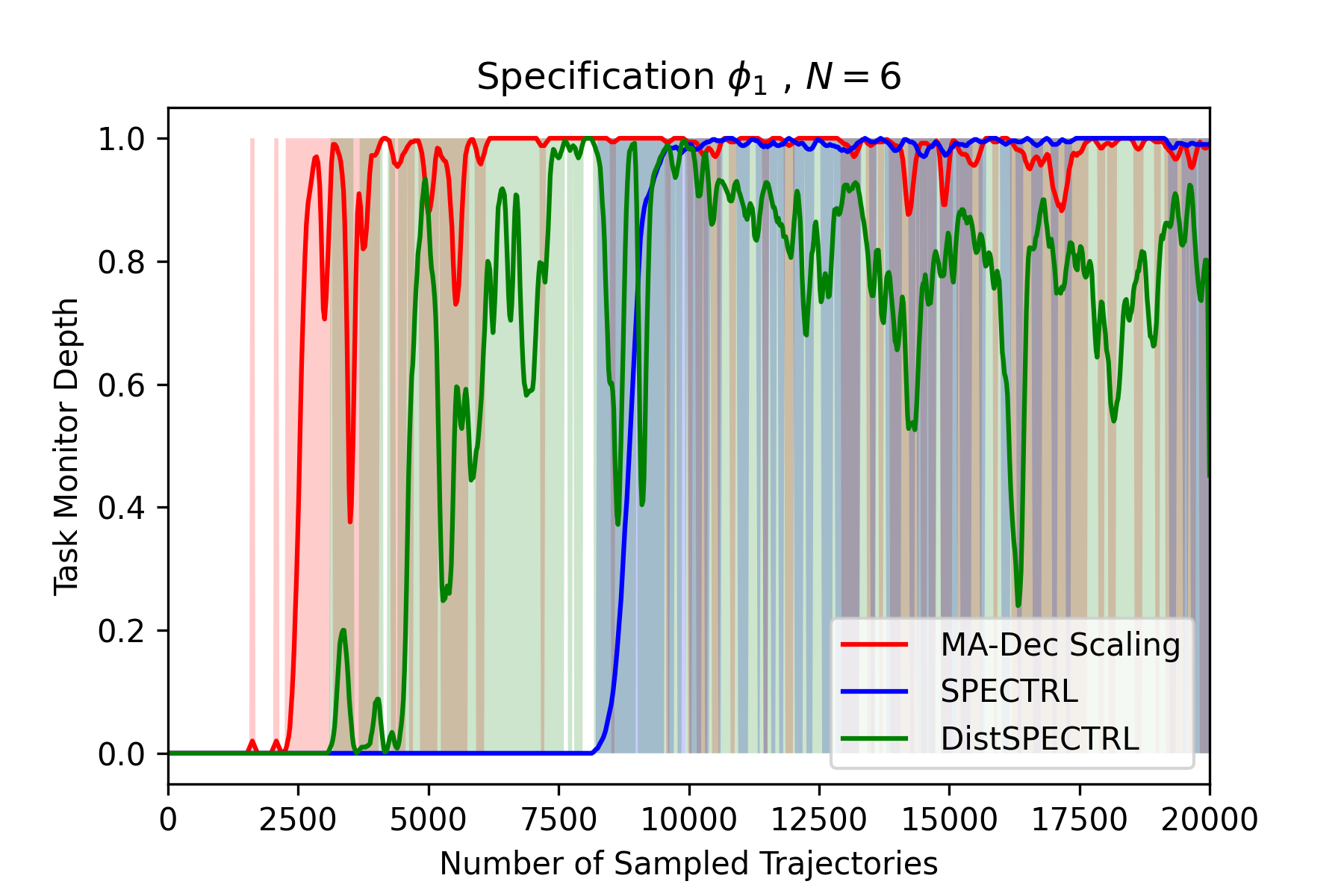}
	\end{center}
	\caption{Specification satisfaction percentages for $N=6$ agents on $\phi_1$ comparing the \scalingname scaling method (red) to centralized \spectrl (blue) and vanilla \toolname (green).}
	\label{fig:phi1_n6_result}
\end{figure}   

%
%
%

\begin{algorithm}[h]
	\SetAlgoLined
	Given Specification $\phi$ that is  MA-Decomposable with factor $k$ and agent set $\mathcal{N}$ of $N$ agents.
	
	Given a function $\texttt{train}(\phi,n_j)$ that trains a policy  satisfying $\phi$ (up to a performance metric) from a previous set of policy parameters for the group of agents $n_j$.

	\SetKwFunction{FMain}{set\_groups}
	\SetKwProg{Fn}{Function}{:}{}
	\Fn{\FMain{$g$, $\mathcal{N}$}}{
		// $g\in \mathbb{Z}^+$ i.e. an integer $g\ge1$
		
		// Makes partition of $\mathcal{N}$ with minimum group size $g$
		
		\uIf{$g>|\mathcal{N}|$}{\KwRet $\{\mathcal{N}\}$}
		\uElse{
			Initialize $j=1, J = \lfloor\frac{N}{g}\rfloor$
			
			\ForEach{$a\in \mathcal{N}$}{
				\uIf{$|n_j| < g\ \mathtt{or}\ j==J$ }{$n_j \leftarrow  n_j\cup {a}$}
				\uElse{
					$j\leftarrow j+1$
					
					$n_j = \{\}$ 
				}
				
			}
			
			\KwRet $\{n_j\}_j$ 
		}
		
	}
	
	Initialize Agent Policies $\Pi$, $i=1$, $g_1 = k$ 
	
	Initialize $\{n_j\}_j=$\FMain{$g_1$, $\mathcal{N}$} 
	
	\While{$|n_1|\le |\mathcal{N}|$}{
		\ForEach{$n_j \in \{n_j\}_j$}{
			Run ${\Pi} \leftarrow \texttt{train}(\phi, n_j)$ independently of agents $\mathcal{N}\setminus n_j$ updating the policies of $n_j$.}
		
		$g_{i+1} = fg_i$.

		\uIf{$g_{i+1} > |\mathcal{N}|$}{
			// Already reached Final Stage with $n_1 == \mathcal{N}$
			
			\KwRet }
		
		$\{n_j\}_j \leftarrow$ \FMain{$g_{i+1}$, $\mathcal{N}$} 
		
		$i\leftarrow i+1$
	}
	\caption{\scalingname Scaling}
	\label{algo:scaling}
\end{algorithm}

\subsection{Calculating Scaling constant $\mathbf{c_k}$}
\label{app:scaling_const}

We want all rewards at stage $i$ to be less than the rewards at stage $i+1$ to prevent local optima from arising where an agent is not incentivized to progress to the next stage.
Assuming that the final reward at all stages is also upper bounded by $C_u$ (as is $\alpha$).

\begin{align*}
r_{i,t} \le	& r_{i+1,t'}  &~\forall t,t'  \\
\implies ic_k + CTM_{i,t} \le& (i+1)c_k + CTM_{i+1,t'} \\
\implies \max(ic_k + CTM_{i,t})\le& \min((i+1)c_k  + CTM_{i+1,t'})\\
\end{align*} 
Since $c_k\in \mathbb{R}$ is a constant we get
\begin{align*}
 \max( CTM_{i,t}) - \min(CTM_{i+1,t'}) \le& c_k\\
\implies C_u - (-2DC_u) \le& c_k
\end{align*} 
Thus a suitable $c_k$ is $(2D+1)C_u$.

\section{Implementation}

\subsection{Computational resources}

All experiments were run on a Intel Xeon Gold 2.10 Ghz 64-core machine with 252 GB of RAM. Individual experiments used no more than 16 cores at a time with experiments involving a hyperparameter search taking 2 cores each.

\subsection{Hyperparameters}
A single 2 layer neural network with 256 nodes each and a $\tanh$ activation function was used. 
The learning rate was varied from $1\times10^{-3}$ to 
$1\times10^{-5}$ over $2\times10^7$ iterations.

We used a grid search on hyperparameters for all experiments. 
\begin{table}[h!]
	\centering
	\captionof{table}{Hyperparameters used for grid search}
	\begin{tabular}{cc}
		\toprule
		Hyperparameter & Ranges  \\
		\midrule
		Batch Size   & $[10000,20000] $   \\
		Initial Learning Rate   & $[10^{-3}, 10^{-4}, 10^{-5}] $   \\
		Entropy Coefficient   & $[0, 0.00176] $ \\
		\bottomrule
	\end{tabular}
	\label{tab:hparams}
	
\end{table}
\subsection{Metrics}
The specification satisfaction is reported with value from
$0$ being no sub-task completed to $1.0$ being the entire
specification satisfied. 
 For more details on the compilation rules we refer readers to Sec. \ref{app:construction}.

\begin{figure}[!h]
	\begin{center}
		\minipage{0.5\linewidth}
	\includegraphics[width=\linewidth, trim={0 0 6cm 0},clip]{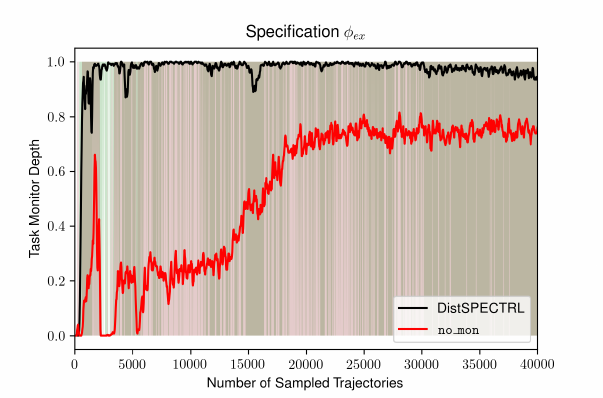}
	\endminipage
	\captionof{figure}{Specification satisfaction percentages for the task monitor shown in Fig. \ref{fig:example_spec} with $
		\phi_{ex}=  \mathtt{reach}_{gl}(10,10); \mathtt{or}\ \left[ \mathtt{reach}_{lo}(3,0) ; \left[ \mathtt{reach}_{lo}(10,10) \ \mathtt{or}\  \mathtt{reach}_{gl}(5,0)   \right] \right]
		$. }
	\label{fig:phi_ex}
\end{center}
\end{figure}

\section{Subtask Synchronization}
\label{app:ss_synch}
\edit{
%
Another example to further strengthen this notion of task synchronization is
\begin{align*}
	\phi_{2a}'=   \left[\mathtt{reach}_{lo}(P) ; \mathtt{reach}_{gl}(Q) \ \right]
	\mathtt{or}\  \left[\mathtt{reach}_{lo}(P') ; \mathtt{reach}_{gl}(Q') \ \right]
\end{align*}
Here, if an agent assumed that it only needed to choose
between the two local objectives $\mathtt{reach}_{lo}(P)$ and
$\mathtt{reach}_{lo}(P')$ and not look further into the future
for the presence of global objectives, a task mismatch would
occur between agents who are in different branches of the task
monitor \textit{i.e.} the sub-specification $\left[\mathtt{reach}_{lo}(P) ;
\mathtt{reach}_{gl}(Q) \ \right]$
vs. $\left[\mathtt{reach}_{lo}(P') ; \mathtt{reach}_{gl}(Q')
\ \right]$.  We see that global objectives deeper in the
sequence of specifications require prior synchronization to
reaching the stage just before task completion.  Thus, we see there is
a marked need for task synchronization among agents as 
specifications become increasingly complex.}

\edit{The identification of local synchronization states
$q\in\synchset$ , where $q\notin Q_g$ is as follows:}
\begin{enumerate}
	\item Select all branching states in the graph of $M_\phi$.
	\item Remove those with all branches local and disconnected.
	
	That is, all monitor states in these branches that only have local transitions $\delta_l$.
	\item Remove all those whose branches rejoin at some state (the rejoin point) and
	have all paths from branching state to the rejoin point not
	include  any global monitor states.
	
	That is, if we consider only the subgraph of  $M_\phi$ starting 
	from the branching state, the  rejoin point should have no ancestors 
	which are global monitor states.
	
\end{enumerate}

\section{Environments}
\subsection{2D Environment}
The environment follows first order dynamics in a 2D space ($\mathcal{S}\in \mathbb{R}^2$). The action space ($\mathcal{A}\in \mathbb{R}^2$) provides the velocity of the agent in the space. Agents are initialized in a line below their reference goals at a Y-coordinate uniformly sampled between $(2,3)$. 

\subsection{3D Environment}
The environment follows first order dynamics in a 3D space ($\mathcal{S}\in \mathbb{R}^3$). The action space ($\mathcal{A}\in \mathbb{R}^3$) provides the velocity of the agent in the space. Agents are initialized below the goal in the X-Y plane and with a random Y,Z coordinate uniformly sampled between $(2,3)$.

\subsection{StarCraft 2}
\label{app:sc2}
\edit{Starcraft 2 \cite{smac_aamas} experiments used 
 the "8m" map with 8 controllable marines and 8 enemy AI-controlled marines. Each agent had state space 
 $\mathcal{S}\in \mathbb{R}^{80}$ and 14 discrete actions.
$\mathtt{away\_from\_enemy}$ defines a predicate that is true when the agent cannot be shot by the enemy.
 $\mathtt{shooting\_range}$ defines a predicate that is true when the agent can shoot the enemy
 (but can also be shot as well).
 $\mathtt{away\_from\_enemy}_{gl}$ being true implies that all the agents are together away from the 
 enemy at once (in a synchronized manner).}

  \edit{To define these predicates we make use of two indicators provided by the Starcraft 2 environment.
  The first being $\mathtt{shooting\_range\_ind} \in \{0,1\}$ which is 1 when the enemy within shooting range and 0 
 otherwise.
 The other is 
  $\mathtt{dist\_to\_enemy} \in [0,1]$ being the normalized distance to an enemy
 which is 0 when the enemy is not visible (the observation radius is larger than the shooting range) .
}

\edit{ The quantitative semantics of these new predicates are then
	\begin{align*}
		\quantpred{\mathtt{away\_from\_enemy}} =& (1-\mathtt{shooting\_range\_ind})*(\epsilon) +\\ &\mathtt{shooting\_range\_ind}*(\mathtt{dist\_to\_enemy} - \epsilon)\\
	\quantpred{\mathtt{shooting\_range}} =& (1-\mathtt{shooting\_range\_ind})*(-\epsilon) +\\ &\mathtt{shooting\_range\_ind}*(\mathtt{dist\_to\_enemy} - \epsilon/10)
\end{align*}
where $\epsilon
\in [0,1]$ is a real-value representing the error tolerance set to 0.5 in the Starcraft experiments.}

\edit{To augment our discrete action space Markov Game for the centralized SPECTRL comparison, we included an additional agent with 
access to the full system state. This centralized controller was used to choose between the available task monitor transitions.
}

\section{Proofs}
\label{app:proofs}
\subsection{Proof of Theorem \ref{mkvgspec-thm}}
The proof follows the exact outline as in \spectrl since the language composition and compilation rules are equivalent in the necessary steps. We repeat their arguments here for clarity. First, the following lemma follows by structural induction:
\begin{lemma}\label{lem:atomic}
	For $\sigma \in \Sigma$, $\pred{\sigma}(s,v)=\true \iff \quantpred{\sigma}(s,v) > 0$.
\end{lemma}

Next, let $G_M$ denote the underlying state transition graph of a task monitor $M$. Then, 
\begin{lemma}\label{lem:monitorproperties}
	The task monitors constructed by our algorithm satisfy the following properties:
	\begin{enumerate}
		\item The only cycles in $G_M$ are self loops. 
		\item The finals states are precisely those states from which there are no outgoing edges except for self loops in $G_M$. 
		\item In $G_M$, every state is reachable from the initial state and for every state there is a final state that is reachable from it.
		\item For any pair of states $q$ and $q'$, there is at most one transition from $q$ to $q'$.
		\item There is a self loop on every state $q$ given by a transition $(q,\top,u,q)$ for some update function $u$ where $\top$ denotes the true predicate.
	\end{enumerate}
\end{lemma}
The first three properties ensure progress when switching from one monitor state to another. The last two properties enable simpler composition of task monitors. The proof follows by structural induction. Theorem \ref{mkvgspec-thm} now follows by structural induction on $\phi$ and Lemmas \ref{lem:atomic} and \ref{lem:monitorproperties}.

\subsection{Proof of Theorem \ref{thm:rewardshaping}}

i) Let $\tilde{\zeta}_m,\tilde{\zeta}_m'$ be two augmented rollouts such that $\tilde{R}^i(\tilde{\zeta}_m) >
\tilde{R}^i(\tilde{\zeta}_m')$. 
\begin{itemize}
	\item  Case A. Both $\tilde{\zeta}_m,\tilde{\zeta}_m'$ end in final monitor states. Here $\tilde{R}^i_s(\tilde{\zeta}_m) = \tilde{R}^i(\tilde{\zeta}_m) >
	\tilde{R}^i(\tilde{\zeta}_m') = \tilde{R}_s^i(\tilde{\zeta}_m')$.
	\item Case B. $\tilde{\zeta}_m$ ends in a final monitor state but $\tilde{\zeta}_m'$ does not. Here

	\begin{align*}
	\tilde{R}_s^i(\tilde{\zeta}_m)=	& \max_{k\le j<T} \alpha(\bar{s}_j,q^i_T,v_j)   \\
	& + 2C_u(d_{q^i_T}-D) + C_l \\
	\le& \max_{k\le j<T} \alpha(\bar{s}_j,q^i_T,v_j)  -2C_u + C_l   &&  (d_{q^i_T}-D\le -1)\\
	\le & C_l && ( C_u\ge\alpha , C_u\ge0)\\
	\le & 	\tilde{R}^i(\tilde{\zeta}_m) && ( C_l\le \tilde{R}^i~\forall~ i\in \mathcal{N})\\
	=& \tilde{R}_s^i(\tilde{\zeta}_m) && ( q^i_T\in F)
	\end{align*} 
	
	\item Case C. $\tilde{\zeta}_m$ ends in a non-final monitor state. Here $\tilde{R}^i(\tilde{\zeta}_m) = -\infty$ and $\tilde{R}^i(\tilde{\zeta}_m') = -\infty$ as well.
	
\end{itemize}

(ii) if
$\tilde{\zeta}_m$ and $\tilde{\zeta}_m'$ end in
distinct non-final monitor states $q^i_T$ and
$(q^i_T)'$ such that $d_{q^i_T} > d_{(q^i_T)'}$, then
$\tilde{R}^i_s(\tilde{\zeta}_m) \geq
\tilde{R}^i_s(\tilde{\zeta}_m')$.

Here the trajectories vary in only one agent's monitor state.
	\begin{align*}
	\tilde{R}_s^i(\tilde{\zeta}_m)=	& \max_{k\le j<T} \alpha(\bar{s}_j,q^i_T,v_j) + C_l  \\
	& + 2C_u(d_{q^i_T}-D)  \\
	\ge& \max_{k\le j<T} \alpha(\bar{s}_j,q^i_T,v_j)  + C_l   &&  (d_{q^i_T}\ge  d_{q^i_T}'+1)\\
	& + 2C_u(  d_{q^i_T}'-D) + 2C_u  \\
	\ge& C_u  + C_l   &&  ( C_u>|\alpha| \implies C_u > -\alpha)\\
	& + 2C_u(  d_{q^i_T}'-D)   \\
	\ge& \max_{k\le j<T} \alpha(\bar{s}_j',(q^i_T)',v_j')  + C_l   &&  (C_u > \alpha)\\
	& + 2C_u(  d_{q^i_T}'-D)   \\
	=& \tilde{R}_s^i(\tilde{\zeta}_m')
	\end{align*} 

\subsection{Proof of Theorem \ref{thm:distributivedecomp}}
\label{proof:distributivedecomp}
Given $\phi$ being MA-Distributive, then for two disjoint sets of agents $n_1, n_2 \subset \mathcal{N}$
$$\phi(\zeta_m,n_1\cup n_2) \implies \phi(\zeta_m,n_1) \land \phi(\zeta_m,n_2) $$
Given a value $k\in \mathbb{Z}^+, 1\le k < N$ we can create a group of agent sets
$\{n_j\}_{j\in{1,\ldots,J}}$ forming a partition of $\mathcal{N}$ with minimum group size $k$ using the 
$\mathtt{set\_groups}(k,\mathcal{N})$ function in Alg.\ref{algo:scaling}.
Now
\begin{align*}
	\phi(\zeta_m,\bigcup_{j\in{1,\ldots,J}}n_j) \implies 	& \phi(\zeta_m,n_1) \land \phi(\zeta_m,\bigcup_{j\in{2,\ldots,J}}n_j)  \\
	\implies & \phi(\zeta_m,n_1) \land \phi(\zeta_m,n_2) \land \phi(\zeta_m,\bigcup_{j\in{3,\ldots,J}}n_j)  \\
	\implies & \bigwedge\limits_{j\in \{1,\ldots,J\}} \phi(\zeta_m,n_j)  
\end{align*} 
Thus $\phi$ is also MA-Decomposable with factors $k\in \mathbb{Z}^+, 1\le k < N$.
